\documentstyle[times,preprint,eqsecnum,aps,epsf]{revtex}
\tightenlines
\newcommand{\be}{\begin{equation}}
\newcommand{\ee}{\end{equation}}
\newcommand{\ba}{\begin{eqnarray}}
\newcommand{\ea}{\end{eqnarray}}
\newcommand{\bra}{\left\langle}
\newcommand{\ket}{\right\rangle}
\newcommand{\sn}{\mathit{sn}}
\newcommand{\cn}{\mathit{cn}}
\newcommand{\dn}{\mathit{dn}}
\begin{document}
\title{\Large{\bf{Relaxation of Wobbling Asteroids and Comets. Theoretical 
Problems. Perspectives of Experimental Observation.}}}
\author{\Large{{Michael Efroimsky}}}
\address{Department of Physics, Harvard University\\} 
\address{~\\}
\address{~\\}
\address{NEW ADDRESS:} 
\address{Institute for Mathematics and Its Applications,
University of Minnesota}
\address{207 Church Street SE, Suite 400, Minneapolis MN 55455 USA\\}
\address{~\\}
\address{e-mail: efroimsk@ima.umn.edu}
\address{telephones: (612) 333 2235 , (612) 625 5532 ; fax: (612) 626 7370}
\author{\\}
\author{PUBLISHED IN:} 
\author{\large{\bf{{{Planetary and Space Science}}, ~Vol. {\bf{49}}, ~p. 937 (2001)}}}
\author{\\}
\author{\\}
\author{\\}
\maketitle

\pagebreak



\begin{abstract}

A body dissipates energy when it freely rotates about any axis different from 
principal. This entails relaxation, i.e., decrease of the rotational energy, 
with the angular momentum preserved. The spin about the major-inertia axis 
corresponds to the minimal kinetic energy, for a fixed angular momentum. Thence
one may expect comets and asteroids (as well as spacecraft or cosmic-dust 
granules) stay in this, so-called principal, state of rotation, unless they are
forced out of this state by a collision, or a tidal interaction, or 
cometary jetting, or by whatever other reason. As is well known, comet 
P/Halley, asteroid 4179 Toutatis, and some other small bodies exhibit very 
complex rotational motions attributed to these objects being in non-principal 
states of spin. Most probably, the asteroid and cometary wobble is quite a 
generic phenomenon. The theory of wobble with internal dissipation has not been
fully developed as yet. In this article we demonstrate that in some spin 
states the effectiveness of the inelastic-dissipation process is several orders
of magnitude higher than believed previously, and can be measured, by the 
presently available observational instruments, within approximately a year 
span. We also show that in some other spin states both the precession and 
precession-relaxation processes slow down considerably. (We call it 
near-separatrix lingering effect.) Such spin states may evolve so slowly that 
they can mimic the principal-rotation state.

\end{abstract}

\pagebreak

\section{Introduction}

It was a surprise for mission experts when, in 1958, the Explorer satellite 
changed its rotation axis. The satellite, a very prolate body with four 
deformable antennas on it, was planned to spin about its least-inertia axis, 
but for some reason refused to do so. Later the reason was understood: on 
general grounds, the body should end up in the spin state that minimises the 
kinetic rotational energy, for a fixed angular momentum. The rotation state 
about the maximal-inertia axis is the one minimising the energy, whereas spin about the 
least-inertia axis corresponds to the maximal energy. As a result, the body 
must get rid of the excessive energy and change the spin axis. This explains 
the vicissitudes of the Explorer mission.

Similarly to spacecraft, a comet or an asteroid in a 
non-principal rotation mode will dissipate energy and will, accordingly, 
return to the stable spin (Black et al. 1999, Efroimsky \& Lazarian 1999). 
Nevertheless, several objects were found in excited 
states of rotation. These are, for example, comet P/Halley (Sagdeev et al. 1989;
Peale \& Lissauer 1989; Peale 1991), comet 46P/Wirtanen (Samarasinha, 
Mueller \& Belton 1996; Rickman \& Jorda 1998), comet 29P/Schwachmann-Wachmann 
1 (Meech et al 1993), asteroid 1620 Geographos (Prokof'eva et al. 1997; 
Prokof'eva et al. 1996), and 
asteroid 4179 Toutatis (Ostro et al. 1993, Harris 1994, Ostro et al. 1995, 
Hudson and Ostro 1995, Scheeres et al. 1998, Ostro et al. 1999).

The dynamics of a freely rotating body is determined, on the one hand, by the 
initial conditions of the object's formation and by the external factors 
forcing the body out of its principal spin state. On the other hand, it is 
influenced by the internal dissipation of the excessive kinetic energy 
associated with wobble. Two mechanisms 
of internal dissipation are known. The so-called Barnett
dissipation, caused by the periodic remagnetisation, is relevant only in the
case of cosmic-dust-granule alignment (Lazarian \& Draine 1997). The other 
mechanism, called inelastic relaxation, is also relevant for mesoscopic 
grains, and plays a primary role in the case of macroscopic bodies

Inelastic relaxation results from the alternating stresses that are 
generated inside a wobbling body by the transversal and centripetal 
acceleration of its parts. The stresses deform the body, and inelastic 
effects cause energy dissipation. 

The external factors capable of driving a rotator into an excited state are 
impacts and tidal interactions, the latter being of a special relevance for 
planet-crossers. In the case of comets, wobble is largely caused by jetting. 
Even gradual outgassing may contribute to the effect because a 
spinning body will start tumbling if it changes its principal axes through a 
partial loss of its mass or through some redistribution thereof. Sometimes the 
entire asteroid or comet may be a wobbling fragment of a progenitor  
disrupted by a collision (Asphaug \& Scheeres 1999, Giblin \& Farinella 1997, 
Giblin et al. 1998) or by tidal forces. All these factors that excite rotators 
compete with the inelastic dissipation that always tends to return the rotator
back to the fold.

Study of rotation of small bodies may provide much information about their 
recent history and internal structure. However, theoretical interpretation of the 
observational data will become possible only after we understand quantitatively 
how inelastic dissipation affects rotation.

Evidently, the kinetic energy of rotation will decrease at a rate 
equal to that of energy losses in the material. Thus one should first 
calculate the elastic energy stored in a tumbling body, and then calculate the
energy-dissipation rate, using the material quality factor $\, Q $. This 
empirical factor is introduced for a phenomenological description of
the overall effect of the various attenuation mechanisms (Nowick \& Berry 
1972; Burns 1986, 1977; Knopoff 1963; Goldreich \& Soter 1965). A comprehensive discussion of the 
$Q$-factor of asteroids and of its frequency- and temperature-dependence is 
presented in Efroimsky \& Lazarian (2000).

A pioneer attempt to study inelastic relaxation in a wobbling asteroid was made
by Prendergast\footnote{We wish to thank Vladislav Sidorenko for drawing our
attention to Prendergast's article, and for reminding us that much that 
appears new is 
well-forgotten past.} back in 1958. Prendergast (1958) pointed out that the 
precession-caused acceleration must create fields of stress and strain over the
body volume. He did not notice the generation of the higher-than-the-second 
harmonics, but he did point out the presence of the second harmonic: 
he noticed that precession with rate $\;\omega\;$ produces stresses of
frequency $\;2\,\omega\;$ along with those of $\;\omega\;$. Since Prendergast's
calculation was wrong in several respects (for example, he missed the term 
$\;{\dot{\bf \Omega}}\,\times\,{\bf r}\;$), it gave him no chance to correctly 
evaluate the role of nonlinearity, i.e., to estimate the relative contribution 
of the harmonics to the entire effect, a contribution that is sometimes of the 
leading order. Nonetheless, Prendergast should be credited for being the first 
to notice the essentially nonlinear character of the inelastic relaxation. 
Another publication where the emergence of the second harmonic was pointed out 
was by Peale (1973) who addressed inelastic relaxation in the case of nearly 
spherical bodies. This 
key observation made by Prendergast and Peale went unnoticed by colleagues and
was forgotten, even though their papers were once in a while mentioned in the 
references. Later studies undertaken by Burns \& Safronov (1973), for 
asteroids, and by 
Purcell (1979), for cosmic-dust granules, treated the issue from scratch and 
fully ignored not only the higher harmonics but even the second mode $\;2\,
\omega\;$. This led them to a several-order-of-magnitude underestimation of the
effectiveness of the process, because the leading effect comes often from the 
second mode and sometimes from higher modes. One more subtlety, missed by everyone
who ever approached this problem, was that, amazingly, the harmonics $\;\omega_n\,=\,
n\,\omega_1\;$ are {\bf{not necessarily multiples of the precession rate}} $\,\bf{\omega}\,$. 
We shall demonstrate that in fact a body precessing at rate $\,\omega\,$ experiences a
superposition of stresses alternating at frequencies $\;\omega_n\,=\,n\,\omega_1\,$. 
Here the ''base frequency''$\,$\footnote{Term suggested by William Newman.} $\,\omega_1\,$ can
either coincide with the precession rate $\,\omega\,$ or be {\bf{lower}} than it, dependent upon the
symmetry of the top and upon its rotation state. (For example, $\,\omega_1\,$ coincides with
$\,\omega\,$ in the case of symmetrical oblate rotator. In this special case, $\,\omega_1\,$ 
itself and $\,\omega_2\,=\,2\,\omega_1\,= \,2\,\omega \,$ are the only emerging harmonics. 
But in the general case of a triaxial top all the other harmonics will show themselves.)

Another oversight present in all  
the afore-quoted studies was their mishandling of the boundary conditions. In 
Purcell's article, where the body was modelled by an oblate rectangular prism, 
the normal stresses had their maximal values on the free surfaces and vanished 
in the centre of the body (instead of being maximal in the centre and vanishing
on the surfaces). In Burns \& Safronov (1973) the boundary conditions, at first
glance, were not touched upon at all. In fact, they were addressed tacitly when
the authors tried to decompose the pattern of deformations into bending and 
bulge flexing. An assumption adopted in Burns \& Safronov (1973), that ``the 
centrifugal bulge and its associated strains wobble back and forth relative to 
the body as the rotation axis {\bf $\; \bf \omega\;$} moves through the body 
during a wobble period,'' lead the authors to a misconclusion that the ``motion
of the bulge through the (nutation) angle $\;\alpha\;$ produces strain 
energy'' and to a calculation based thereon. In reality, however, the bulge 
appearance is but an iceberg tip, in that an overwhelming part of the inelastic
dissipation process is taking place not near the surface but in the depth of 
the body, deep beneath the bulge. This follows from the fact that the 
stress and strain are small in the shallow regions and increase in the depth, 
if the boundary conditions are improved. (Remarkably, in the Peale, Cassen and 
Reynolds (1979) discussion of tidal dissipation in Io, maximum dissipation was 
also found to occur in the centre of the initially solid Io.)

\section{Notations and Assumptions}

We shall use two coordinate systems. The body frame will naturally be  
represented by the three principal axes of inertia: $1$, $2$, and $3$, with 
coordinates $x$, $y$, $z$, and unit vectors ${\bf{e}}_{1}$, 
${\bf{e}}_{2}$, ${\bf{e}}_{3}$. The second (inertial) frame ($X$, $Y$, 
$Z$), with basis vectors ${\bf{e}}_{X}$, ${\bf{e}}_{Y}$, 
${\bf{e}}_{Z}$, may be chosen with its $Z$ axis aimed along the body's
angular-momentum vector $\bf{J}$ and with its origin coinciding 
with that of the body frame (i.e., with the centre of mass). The inertial-frame
coordinates will be denoted by the same capital letters: $X$, $Y$, and $Z$ as 
the axes.

The angular momentum  $\bf{J}$ of a freely-precessing body is  
conserved in the inertial frame. The {\it 
{body-frame-related}} components of the inertial angular velocity $\bf{\Omega}$
will be called $\; \Omega_{1,2,3} \;$, while letter $\;\omega\;$ will be 
reserved for the precession rate. 

A body-frame-based observer will view both the inertial angular 
velocity $\;{\bf{\Omega}}\;$ and the angular momentum $\;\bf J\;$ nutating
around the principal axis $\;3\;$ at rate $\;\omega\,$. An inertial observer, 
though, will argue that it is rather axis $3$ and angular velocity $\;\bf 
\Omega\;$ that are wobbling about $\bf{J}$. As is well known, the precession 
rate of $\;\bf \Omega\;$ about $\;\bf J\;$ in the inertial frame is different 
from the precession rate $\;\omega\;$ of  $\;\bf \Omega\;$ about axis $\;3\;$ 
in the body frame. (See Section IV.) We would emphasize that the precession 
rate of our interest is the one in the body frame.

Free rotation of a body is described by Euler's equations
\be
\frac{d}{dt} \; \left(I_i \; \Omega_i \protect\right) \; = \; \left( 
I_j \; - \; I_k \protect\right) \; \Omega_j \; \Omega_k\;\;\;,\;\;\;
\label{2.1}
\ee
$(ijk)\,$ being a cyclic transposition of $\,(123)$, and the principal 
moments of inertia ranging as
\be
I_1 \; \leq \;   I_2 \; \leq \;   I_3 \; \; \; \; .
\label{2.2}
\ee
Since no uniform concensus on notations exists in the literature, the following
table may simplify reading: 
\nopagebreak
\begin{table}
\begin{tabular}{|p{3cm}|p{3cm}|p{3cm}|p{3cm}|}
\hline
         &                  &                             &                 \\
         &Principal moments & Components of the angular   & Rate of angular-velocity wobble,  \\
         &of inertia        & velocity in the body frame  & in the body frame\\
         &                  &                             &             \\
\hline
                     &                             &                    &\\
  (Purcell 1979),    &                             &                    &\\
  (Lazarian \&       &                             &                    &\\ 
  Efroimsky 1999),   &                             &                    &\\
  (Efroimsky 2000),  &                             &                    &\\ 
  (Efroimsky \&      & $I_3\;\geq\;I_2\;\geq\;I_1$  &$\Omega_3\;\;,\;\;\;\Omega_2\;\;,\;\;\;\Omega_1$&
$\omega$\\
  Lazarian 2000),    &                             &                    &\\ 
  present article    &                             &                    &\\
                     &                             &                    &\\
\hline
          &                 &                                           &\\
(Synge \& Griffiths 1959) & $A\geq B\geq C$ & $\omega_1\;\;,\;\;\;\omega_2\;\;,\;\;\;\omega_3$ &$p$\\
 & & &\\
\hline
 & & &\\
(Black et al. 1999) & $C\geq B\geq A$ & $\omega_c\;\;,\;\;\;\omega_b\;\;,\;\;\;\omega_a$ & $\nu$\\
 & & &\\
\hline
\end{tabular}
\end{table}
In the body frame, the period of angular-velocity precession about the principal axis $\;3\;$ is: 
$\;\tau\;=\;2\,\pi/\omega\,.$ Evidently, 
\begin{eqnarray}
{\dot{\Omega}}_i/{\Omega}_i \; \approx \; {\tau}^{-1} \; \; \; , 
\; \; \; \; \; 
{\dot{I}}_i/{I}_i \; \approx \; {\tau}^{-1} \, \epsilon \; \; \;,\;\;\;
\label{2.3}
\end{eqnarray}
$\epsilon\;$ being a typical value of the relative strain that is several 
orders less than unity. These estimates lead to the inequality  $\; \dot{I_i}\,
\Omega_i\;\ll \;I_i\, \dot{\Omega_i}\;$, thereby justifying the commonly used 
approximation to Euler's equations:
\be
I_i \; {\dot{\Omega}}_i \; = \; \left(I_j \; - \; I_k \protect\right) \; 
\Omega_j \; \Omega_k \; \; \; \; .
\label{2.4}
\ee
Thus it turns out that in our treatment the same phenomenon is neglected in one
context and accounted for in another: on the one hand, the very process of the 
inelastic dissipation stems from the precession-inflicted small deformations;  
on the other hand, we neglect these deformations in order to write down
(\ref{2.4}). This approximation (also discussed in Lambeck 1988) may be called
adiabatic, and it remains acceptable insofar as the relaxation is slow against 
rotation and precession. To cast the adiabatic approximation into its exact 
form, one should first come up with a measure of the relaxation rate. Clearly,
this should be the time derivative of the angle $\;\theta\;$ made by the 
major-inertia axis $\;3\;$ and the angular momentum $\;\bf J\;$. The axis 
aligns towards $\;\bf J\;$, so $\;\theta\;$ must eventually decrease. Be 
mindful, though, that even in the absence of dissipation, $\;\theta\;$ does 
evolve in time, as can be shown from the equations of motion. Fortunately, this
evolution is periodic, so one may deal with a time derivative of the angle 
averaged over the precession period. In practice, it turns out to be more 
convenient to deal with the squared sine of $\theta$ (Efroimsky 2000) and 
to write the adiabaticity assertion as:
\be
-\;\frac{d\,\bra \sin^2 \theta \ket}{dt}\,\ll\,\omega\;\;,
\label{2.5}
\ee
$\omega$ being the precession rate and $\,<...>\,$ being the average  
over the precession period. The case of an oblate symmetrical 
top\footnote{Hereafter oblate symmetry will imply not a 
geometrical symmetry but only the so-called dynamical symmetry: $\,I_1\,=\,
I_2$.} is exceptional, in that $\,\theta \,$ remains, when dissipation is 
neglected, constant over a precession cycle. No averaging is needed, and 
the adiabaticity condition simplifies to:
\be
-\;\left(\frac{d\,\theta}{dt}\right)_{(oblate)}\,\ll\,\omega\;\;.
\label{2.6}
\ee
We would emphasise once again that the distinction between the oblate and triaxial 
cases, distinction resulting in
the different forms of the adiabaticity condition, stems from the difference in
the evolution of $\,\theta \,$ in the weak-dissipation limit. The equations of 
motion of an oblate rotator show that, in the said limit, $\,\theta \,$ stays 
virtually unchanged through a precession cycle (see section IV below). So the 
slow decrease of $\,\theta \,$, accumulated over many periods, becomes an  
adequate measure for the relaxation rate. The rate remains slow, compared to 
the rotation and precession, insofar as (\ref{2.6}) holds. In 
the general case of a triaxial top the equations of motion show that, even in 
the absence of dissipation, angle $\,\theta \,$ periodically evolves, though 
its average over a cycle stays unchanged (virtually unchanged, when 
dissipation is present but weak)\footnote{See formulae (A1) - (A4) in the 
Appendix to Efroimsky 2000.}. In this case we should measure the
relaxation rate by the accumulated, over many cycles, change in the average of
$\,\theta\,$ (or of $\,\sin^2 \theta\,$). Then our assumption about the 
relaxation being slow yields (\ref{2.5})

The above conditions (\ref{2.5}) - (\ref{2.6}) foreshadow the applicability 
domain of our further analysis. For example, of the two quantities,
\be
I_1^2 \, {\Omega}_1^2 \; + \; I_2^2 \, {\Omega}_2^2 \; + \; I_3^2 \, 
{\Omega}_3^2 \; = \; {\bf{J}}^2 \; \; \; , \;\;
\label{2.7}
\ee
\be
I_1 \, {\Omega}_1^2 \; + \; I_2 \, {\Omega}_2^2 \; + \; I_3 \, 
{\Omega}_3^2 \; = \; 2 \; T_{\small{kin}}\;\;\;,\;\; 
\label{2.8}
\ee
only the former will conserve exactly, while the latter will remain virtually 
unchanged through one cycle and will be gradually changing through many cycles 
(just like $\;\;\bra \sin^2 \theta \ket\;\,$).

\section{The strategy}

As mentioned above, in the case of an oblate body, when the moments
of inertia relate as $\;I_3\;>\;I_2\;=\;I_1\;$, the angle $\;\theta\;$ between 
axis 3 and $\;\bf J\;$ remains adiabatically unchanged over the precession 
cycle. Hence in this case we shall be interested in $\dot{\theta}$, the 
rate of the maximum-inertia axis' approach to the direction of $\bf{J}$. In the
general case of a triaxial rotator, angle $\;\theta\;$ evolves through the 
cycle, but its evolution is almost periodic and, thus, its average over 
the cycle remains virtually constant. Practically, it will turn out to be
more convenient to use the average of its squared sine. In this case, the 
alignment rate will be characterised by the time derivative of $\;\;<\sin^2 
\theta>\;$. Evidently,  
\be
\frac{d\;<\sin^2 \theta >}{dt}\;=\;\frac{d\;<\sin^2\theta>}{dT_{kin}}\;\;
\frac{dT_{kin}}{dt}\;\;\;,\;\;\;\;
\label{3.1}
\ee
while for an oblate case, when $\,\theta\,$ remains virtually unchanged over a 
cycle, one would simply write:
\be
\left(\frac{d\, \theta }{dt}\right)_{(oblate)}\;=\;\left(\frac{d\,\theta}{
dT_{kin}}\right)_{(oblate)}\;\;\frac{dT_{kin}}{dt}\;\;\;.\;\;\;\;
\label{3.2}
\ee
The derivative $\;d\,<\sin^2\theta>/dT_{kin}\,$ appearing in (\ref{3.1}), as 
well as $\,\left(d\,\theta/dT_{kin}\right)_{(oblate)}\,$ appearing in 
(\ref{3.2}), can be calculated from (\ref{2.7}), (\ref{2.8}) and the equations
of motion. These derivatives indicate how the losses of the rotational energy 
affect the value of $\;<\sin^2 \theta>\;$ (or simply of $\theta$, in the 
oblate case). The kinetic-energy decrease is caused by the inelastic 
dissipation,  
\be
\dot{T}_{kin} \; = \; < \dot{W} > \;\;\;,\;\;\;
\label{3.3}
\ee
$W\;$ being the elastic energy of the alternating stresses, and $\;<W>\;$ being
its average over a precession cycle. This averaging is justified within our 
adiabatic approach. So we shall 
eventually deal with the following formulae for the alignment rate:
\be
\frac{d\,<\sin^2\theta>}{dt}\;=\;\frac{d\,<\sin^2\theta>}{dT_{kin}}\;\;
\frac{d\,<W>}{dt}\;\;\;,\;\;\;
\label{3.4}
\ee
in the general case, and
\be
\left(\frac{d\, \theta }{dt}\right)_{(oblate)}\;=\;\left(\frac{d\,\theta}{
dT_{kin}}\right)_{(oblate)}\;\;\frac{d\,<W>}{dt}\;\;\;,\;\;\;\;
\label{3.5}
\ee
for an oblate rotator.

Now we are prepared to set out the strategy of our further work. While 
calculation of $\;d {\small \left\langle \right.} \sin^2 \theta {\small {\left. \ket}} / 
dT_{kin}\;$ and $\;\left(d\theta /dT_{kin}\right)_{oblate}\;$ is an easy exercise\footnote{See 
formula (\ref{5.18}) below and also
formulae (A12 - A13) in Efroimsky 2000.}, our main goal will be to find the dissipation rate 
$\;d\,<W>/dt\;$. This quantity will consist of inputs from the dissipation
rates at all the frequencies involved in the process, i.e., from the harmonics at 
which stresses oscillate in a body precessing at a given rate $\,\omega\,$. The stress 
is a tensorial extension of the notion of a pressure or force. Stresses naturally 
emerge in a spinning body due to the centripetal and transversal accelerations of 
its parts. Due to the precession, these stresses contain time-dependent components. 
If we find a solution to the boundary-value problem for alternating stresses, it 
will enable us to write down explicitly the time-dependent part of the elastic energy 
stored in the wobbling body, and to separate contributions from different harmonics:
\be
<W>\;=\;\sum_{n}\;\,<W(\omega_n)>\;\;\;\;\;.\;\;
\label{4.9}
\ee
$W(\omega_n)\;$ being the elastic energy of stresses alternating at frequency 
$\,\omega_n$. One should know each contribution $W(\omega_n)$, for these will determine
the dissipation rate at the appropriate frequency, through the 
frequency-dependent empirical quality factors. The knowledge of these factors, along with
the averages $\,<W(\omega_n)>\,$, will enable us to find the dissipation rates at each harmonic.
Sum of those will give the entire dissipation rate due to the alternating stresses emerging in 
a precessing body.

\section{Inelastic Dissipation}

Equation $(\ref{4.9})$ implements the most important observation upon which all our
study rests: generation of harmonics in the stresses inside a precessing rigid body.
The harmonics emerge because the acceleration of a point inside a precessing body 
contains centrifugal terms that are quadratic in the angular velocity $\bf{\Omega}$. In 
the simpliest case of a symmetrical oblate body, for example, the body-frame-related 
components of the angular velocity are given in terms of $\,\sin \omega t\,$ and $\,\cos 
\omega t\,$ (see formulae (\ref{5.2}) from Section V). Evidently, squaring of $\bf{\Omega}$ 
will yield terms both with $\,\sin \omega t\,$ or $\,\cos \omega t\,$ and with $\,\sin 2 
\omega t\,$ or $\,\cos 2 \omega t\,$. The stresses produced by this acceleration will, too,
contain terms with frequency $\,\omega t\,$ as well as those with the harmonic $\,2\omega t$. 
In the further sections we shall explain that a triaxial body precessing at rate $\,\omega\,$ 
is subject, in distinction from a symmetrical oblate body, to a superposition of stresses 
oscillating at frequencies $\;\omega_n\,=\,n\,\omega_1\,$, the ''base frequency'' $\,\omega_1\,$ 
being lower than the precession rate $\,\omega$. The basic idea is that in the general, 
non-oblate case, the time dependence of the acceleration and stresses will be expressed not 
by trigonometric but by elliptic functions whose expansions over the trigonometric functions 
will generate an infinite number of harmonics.

The total dissipation rate will be a sum of the particular rates (Stacey 1992) 
to be calculated empirically. The empirical description of attenuation is
based on the quality factor $\,Q(\omega)\,$ and on the assumption of 
attenuation rates at different harmonics being independent from one 
another:
\be
\dot{W}\;=\;\sum_{n} \; \dot{W}{({\omega_n})}\;=\;-\;\sum_{n}\;
\frac{\omega_n\;W_0({\omega}_n)}{Q({\omega_n})}\;=\;-\;2\;\sum_{n}\;
\frac{\omega_n\;\,<W({\omega}_n)>}{Q({\omega_n})}\;\;\;\;\\
\label{4.10}
\ee
$\;Q(\omega)\;$ being the quality factor of the material, and  
$\;W_0({\omega}_n)\;$ and $\;\,<W({\omega}_n)>\,\;$ being the maximal and 
the average values of the appropriate-to-$\omega_n\;$ fraction of elastic 
energy stored in the body. This expression will become more general if we put 
the quality factor under the integral, implying its possible coordinate 
dependence\footnote{In strongly inhomogeneous precessing bodies the 
attenuation may depend on location.}:
\be
\dot{W}\;=\;-\;2\;\sum_{\omega_n}\;\int\;dV\;\left\{\frac{\omega_n}{
Q({\omega_n})}\;\,\frac{d\,<W(\omega_n)>}{dV}\;\right\}\;\;\;,\;\;
\label{4.11}
\ee
The above assumption of attenuation rates at different harmonics being mutually 
independent is justified by the extreme smallness of strains (typically, much less 
than $\;10^{-6}$) and by the frequencies being extremely low ($10^{-5}\,-\,10^{-3}\;
Hz$). One, thus, may say that the problem is highly nonlinear, in that we shall take 
into account the higher harmonics in the expression for stresses. At the same time, 
the problem remains linear in the sense that we shall neglect any nonlinearity stemming 
from the material properties (in other words, we shall assume that the strains are linear 
function of stresses). We would emphasize, though, that the nonlinearity is most 
essential, i.e., that the harmonics $\;\omega_n\;$ come to life unavoidably: no matter 
what the properties of the material are, the harmonics do emerge in the expressions for 
stresses. Moreover, as we shall see, the harmonics interfere with one another due to $\;
W\;$ being quadratic in stresses. Generally, all the infinite amount of multiples of $\;
\omega_1\;$ will emerge. The oblate case, where only $\;\omega_1\;$ and $\;2\omega_1\;$ 
show themselves, is an exception. Another exception is the narrow-cone precession of a
triaxial rotator studied in Efroimsky (2000): in the narrow-cone case, only the first
and second modes are relevant (and $\,\omega_1\,\approx \,\omega$). 

Often the overall dissipation rate, and therefore the relaxation rate is determined 
mostly by harmonics rather than by the principal frequency. This fact was discovered 
only recently (Efroimsky \& Lazarian 2000, Efroimsky 2000, Lazarian \& Efroimsky 1999), and 
it led to a considerable re-evaluation of the effectiveness of the 
inelastic-dissipation mechanism. In some of the preceding publications, its 
effectiveness had been underestimated by several orders of magnitude, and the main 
reason for this underestimation was neglection of the second and higher harmonics. As
for the choice of values of the quality factor $\;Q\,$, Prendergast (1958) and
Burns \& Safronov (1973) borrowed the terrestial seismological data for $Q$. 
In Efroimsky \& Lazarian (2000), we argue that these data are inapplicable to 
asteroids.

To calculate the afore mentioned average energies $\,<W(\omega_n)>\,$, we use 
such entities as stress and strain. As already mentioned above, the stress is a tensorial 
generalisation of the notion of pressure. The strain tensor is analogous to the stretching 
of a spring (rendered in dimensionless fashion by relating the displacement to the base 
length). Each tensor component of the stress consists of two inputs, elastic and plastic. 
The former is related to the strain through the elasticity constants of the material; the 
latter is related to the time-derivative of the strain, through the viscosity coefficients.
As our analysis is aimed at extremely small deformations of cold bodies, the 
viscosity may well be neglected, and the stress tensor will be approximated, to a 
high accuracy, by its elastic part. Thence, according to Landau \& Lifshitz (1976), the 
components of the elastic stress tensor $\sigma_{\it{ij}}$ are interconnected 
with those of the strain tensor $\epsilon_{\it{ij}}$ like:
\be
\epsilon_{ij} \; \; = \; \; \delta_{ij} \; \; \frac{Tr \; 
\sigma}{9 \; K} \; \; + \; \; 
\left( \; \sigma_{\it{ij}} \; \; - \; \; \frac{1}{3} \; \; \delta_{ij} 
\; \; Tr \; \sigma \right) \; \frac{1}{2 \; \mu} \; \; \; ~~~,
\label{4.6}
\ee
$\mu$ and $K$ being the {\it{adiabatic}} shear and bulk moduli, and $Tr$ 
standing for the trace of a tensor. 

To simplify the derivation of the stress tensor, the body will be modelled  by 
a rectangular prism of dimensions $\,2\,a\,\times\,2\,b\,\times\,2\,c\,$ where 
$\,a\,\ge\,b\,\ge\,c$. The tensor is symmetrical and is defined by 
\be
\partial_{i}\sigma_{ij}\;=\;\rho\;a_j\;\;,\;\;
\label{4.12}
\ee
$a_j$ being the time-dependent parts of the acceleration components, and 
$\,\rho\,a_j$ being the time-dependent parts of the components of the 
force acting on a unit volume\footnote{Needless to say, these acceleration 
components $\,a_j\,$ are not to be mixed with $\,a\,$ which is the longest 
dimension of the prism.}. Besides, the tensor $\,\sigma_{ij}$ must obey the 
boundary conditions: its product by normal unit vector, $\,\sigma_{ij}n_j\,$, 
must vanish on the boundaries of the body (this condition was 
not fulfilled in Purcell (1979)). 

Solution to the boundary-value problem provides such a distribution of the 
stresses and strains over the body volume that an overwhelming share of 
dissipation is taking place not near the surface but in the depth of the body. 
For this reason, the prism model gives a good approximation to realistic 
bodies. Still, in further studies it will be good to generalise our solution to
ellipsoidal shapes. Such a generalisation seems to be quite achievable, judging
by the recent progress on the appropriate boundary-value problem (Denisov \& 
Novikov 1987).

Equation (\ref{4.12}) has a simple scalar analogue\footnote{This very simple and, 
nonetheless, so illustrative example was kindly offered to me by William Newman.} 
Consider a non-rotating homogeneous liquid planet of radius $\,R\,$ and density 
$\,\rho\,$. Let $\;g(r)\;$ and $\,P(r)\,$ be the free-fall acceleration and the 
self-gravitational pressure at the distance $\,r\,\le \,R\,$ from the centre. 
(Evidently, $\,g(r)\,=\,(4/3) \, \pi \, G \, \rho \, r\,$.) Then the analogue to 
(\ref{4.12}) will read:
\be
\rho \; g(r) \;=\;-\;\frac{\partial P(r)}{ \partial r} \;\;\;\;\;\;,\;\;\;\;
\label{444}
\ee
the expression $\;\rho \, g(r)\;$ standing for the gravity force acting upon a unit volume, 
and the boundary condition being $\;P(R)\,=\,0$. Solving equation (\ref{444}) reveals that the 
pressure has a maximum at the centre of the planet, although the force is greatest at the surface. 
Evidently, the maximal deformations (strains) also will be experienced by the material near 
the centre of the planet. 

In our case, the acceleration $\,\bf a\,$ of a point inside the precessing body will be given not by 
the free-fall 
acceleration $\,g({\bf{r}})\,$ but will be a sum of the centripetal and transversal accelerations: $\,
{\bf{\Omega}}\,\times\,( {\bf{\Omega}}\,\times\,{\bf{r}} )\,+\,{\bf{\dot{\Omega}}}\,\times\,{\bf{r}}\;$, 
the Coriolis term being negligibly small. Thereby, the absolute value of $\,{\bf{a}}\,$ will be proportional 
to that of $\,\bf r\,$, much like in the above example. In distinction from the example, though, the 
acceleration of a point inside a wobbling top will have both a constant and a periodic component, the 
latter emerging due to the precession. For example, in the case of a symmetrical oblate rotator, the 
precessing components of the angular velocity $\,\bf \Omega\,$ will be proportional to $\,\sin \omega 
t\,$ and $\,\cos \omega t\,$, whence the transversal acceleration will contain frequency $\,\omega\,$ 
while the centripetal one will contain $\,2 \omega$. The stresses obtained through (\ref{4.12}) will 
oscillate at the same frequencies, and so will the strains. As we already mentioned, in the case of 
a non-symmetrical top an infinite amount of harmonics will emerge, though these will be obertones not 
of the precession rate $\,\omega\,$ but of some different ''base frequency'' $\,\omega_1\,$ that is
less than $\,\omega.$

Here follows the expression for the (averaged over a precession period) elastic energy stored in a unit 
volume of the body:
\ba
\frac{d\;\,<W>}{dV}\;=\;\frac{1}{2}\;\,<\epsilon_{ij}\;\sigma_{ij}>\;=
\;\frac{1}{4 \mu}\; 
\left\{ \left(\frac{2 \; \mu}{9\; K} \; - \; \frac{1}{3} \protect\right) \;
\,<\,\left(Tr\;\sigma \protect\right)^2\,>\;+\;<\sigma_{ij}\,\sigma_{ij}> 
\protect\right\} \; = 
\nonumber \\
\nonumber \\
\frac{1}{4\mu}\;\left\{\,-\,\frac{1}{1\,+\,\nu^{-1}}\;\,<\left(Tr\;\sigma
\protect\right)^2>\,+\,<\sigma_{xx}^2>\,+\,<\sigma_{yy}^2>\,+\,<\sigma_{zz}^2
> \, + \, 2 \,<\,\sigma_{xy}^2 \, + \, \sigma_{yz}^2 \, + \, 
\sigma_{zx}^2 \,> \protect\right\}\;\;
\label{4.7}
\ea 
where $2\mu/(9K)-1/3= -\nu/(1+\nu) \approx -1/5$, $\,\nu$ being
Poisson's ratio (for most solids $\,\nu \approx 1/4$). Naturally\footnote{Very naturally indeed, 
because, for example, $\,\sigma_{xx} \epsilon_{xx} dV\,=\,(\sigma_{xx}\,dy\,dz)(\epsilon_{xx}\,dx)\,$
is a product of the $\,x$-directed pressure upon the $\,x$-directed elongation of the elementary
volume $dV$.}, the total averaged 
elastic energy is given by the integral over the body's volume: 
\be
<W>\;=\;\frac{1}{2}\;\int\;dV\;\sigma_{ij}\;\epsilon_{ij} \;\;\;\;,\;\;\;
\label{4.8}
\ee
and it must be expanded into the sum (\ref{4.9}) of inputs from oscillations of 
stresses at different frequencies. Each term $\,\bra W(\omega_n) \ket\,$ emerging 
in that sum will then be plugged into the expression (\ref{4.10}), together with the 
value of $\,Q\,$ appropriate to the overtone $\,\omega_n$.

\section{A Special Case: Precession of an Oblate Body.}

An oblate body has moments of inertia that relate as:
\be
I_3\;>\;I_2\;=\;I_1\;\equiv\;I\;\;.\;\;\;\;\;\;
\label{5.1}
\ee
We shall be interested in $\dot{\theta}$, the rate of the maximum-inertia axis'
approach to the direction of angular momentum $\bf{J}$. To achieve this goal, 
we shall have to know the rate of energy losses caused by the periodic 
deformation. To calculate this deformation, it will be necessary to find the 
acceleration experienced by a particle located inside the body at a point ($x$,
$y$, $z$). Note that we address the inertial acceleration, i.e., the one with 
respect to the inertial frame $(X,Y,Z)$, but we express it in terms of 
coordinates $x$, $y$ and $z$ of the body frame $(1,2,3)$ because eventually we
shall have to compute the elastic energy stored in the entire body (through 
integration of the elastic-energy density over the body volume). 

The fast motions (revolution and precession) obey, in the adiabatical 
approximation,  the simplified Euler equations (\ref{2.4}). Their solution,  
with neglect of the slow relaxation, looks (Fowles and Cassiday 1986, Landau and Lifshitz 1976), in the oblate case (\ref{5.1}):
\be
{\Omega}_1 \; \; = \; \; {\Omega}_{\perp} \cos {\omega}t~~,~~~
{\Omega}_2 \; \; = \; \; {\Omega}_{\perp} \sin {\omega}t~~,~~~
{\Omega}_3 \; \; = \; \; const
\label{5.2}
\ee
where
\be
{{\Omega}_{\perp}} \; \; \equiv \; \; {\Omega} \; \; \sin \; {\alpha}~~, ~~~~~
{{\Omega}_{3}} \; \; \equiv \; \; {\Omega} \; \; \cos \; {\alpha}\;\;\;,
\label{5.3}  
\ee
$\;\alpha\;$ being the angle made by the major-inertia axis 3 with $\,\bf{
\Omega}\,$. Expressions (\ref{5.2}) show that in the body frame the angular 
velocity ${\bf{\Omega}}$ describes a circular cone about the principal axis
$3$ at a constant rate 
\be
\omega \; = \; (h - 1) \Omega_3, 
\; \; \; \; \; \; \; \;  h\equiv {I_3}/I~~~. 
\label{5.4}
\ee
So angle $\;\alpha\;$ remains virtually unchanged through a cycle (though in the
presence of dissipation it still may change gradually over many cycles). 
The precession rate $\;\omega \;$ is of the same order as $|{\bf{\Omega}}|$, 
except in the case of $\;h\,\rightarrow\,1\;$ or in a very special case of 
${\bf{\Omega}}$ and ${\bf{J}}$ being orthogonal or almost orthogonal to the 
maximal-inertia axis $3$. Hence one may call not only the rotation but also the
precession ``fast motions'' (implying that the relaxation process is a slow 
one). Now, let $\;\theta\;$ be the angle between the principal axis $3$ and the
angular-momentum $\;\bf J\;$, so that $\;J_3\,=\,J\,\cos \theta\;$ and
\be
{\Omega}_3 \; \; \equiv \; \; \frac{J_3}{I_3} \; \; = \; \;\frac{J}{I_3} \; \; 
\cos \; \theta~~ ~~~~
\label{5.5}
\ee
wherefrom
\be
\omega\;=\;(h\;-\;1)\;\frac{J}{I_3}\;\cos \theta\;\;\;.
\label{5.6}
\ee
Since, for an oblate object,
\be 
{\bf J}\,=\,I_1\,\Omega_1\,{\bf e}_1\,+\,I_2\,\Omega_2\,{\bf e}_2\,+\,I_3\,
\Omega_3\,{\bf e}_3\,=\,I\,(\Omega_1\,{\bf e}_1\,+\,\Omega_2\,{\bf e}_2)\,+\,
I_3\,\Omega_3\,{\bf e}_3\,\;,\;
\label{5.7}
\ee
the quantity $\;\Omega_{\perp}\,\equiv\,\sqrt{\Omega_1^2+\Omega_2^2}\;$ is 
connected with the absolute value of $\bf J$ like:
\be
{\Omega}_{\perp} \; \; = \;\;\frac{J}{I}\;\;\sin\;\theta\;\;=\; \; \frac{J}{I_3} \; h \; \; \sin \; \theta\;\;\;,\;\;\;\;\;\;h\;\equiv\;I_3/I\;\;.\;\;
\label{5.8}
\ee
It ensues from (\ref{5.3}) that $\;{{\Omega}_{\perp}}/{{\Omega}_{3}}\;=\; 
\tan \; \alpha \;$. On the other hand, (\ref{5.5}) and (\ref{5.8}) entail: $\;
{{\Omega}_{\perp}}/{{\Omega}_{3}}\,=\,h\,\tan \, \theta\;$. Hence,
\be
\tan \alpha\;=\;h\;\tan \theta\;\;\;.\;\;\;
\label{5.9}
\ee
We see that angle $\;\theta\;$ is almost constant too (though it gradually 
changes through many cycles). We also see 
from (\ref{5.7}) that in the body frame the angular-momentum vector $\;\bf J\;$
describes a circular cone about axis $3$ with the same rate $\;\omega\;$ as 
$\;\Omega\;$. An inertial observer, though, will insist that it is rather axis 
$3$, as well as the angular velocity $\;\Omega\;$, that is describing circular
cones around $\bf{J}$. It follows trivially from (\ref{5.4}) and (\ref{5.7}) that
\be
I\;{\bf \Omega}\;=\;{\bf J}\;-\;I\;\omega\;{\bf e}_3 \;\;\;,\;\;\;\;
\label{5.10}
\ee
whence it is obvious that, in the inertial frame, both $\bf \Omega$ and 
axis $\;3\;$ are precessing about $\bf J$ at rate $J/I$. (The angular 
velocity of this precession is ${\dot{\bf e}}_3\,=\,{\bf \Omega} \,\times\,{
\bf e}_3\,=\,({\bf J}/I\,-\,\omega\,{\bf e}_3)\,\times\,{\bf e}_3\,=\,({\bf 
J}/I)\,\times\,{\bf e}_3$.) Interestingly, the rate $\omega\,=\,(h\,
-\,1)\,\Omega_3$, at which $\bf \Omega$ and $\bf J$ are precessing 
about axis $\;3\;$ in the body frame, differs considerably from the rate $
J/I$ at which $\bf \Omega$ and axis $\;3\;$ are precessing around $\bf J$ in 
the inertial frame. (In the case of the Earth, $J/I\,\approx\,400\,
\omega\;$ because $\,h\,$ is close to unity.) Remarkably, the 
inertial-frame-related precession rate is energy-independent and, thus, stays 
unchanged through the relaxation process. This is not the case for the 
body-frame-related rate $\,\omega\,$ which, according to (\ref{5.6}), gradually
changes because so does $\,\theta$. 

As is explained above, we shall 
be interested in the body-frame-related components $\Omega_{1,2,3}$ precessing 
at rate $\omega$ about the principal axis $\;3$. Acceleration of an arbitrary 
point of the body can be expressed in terms of these components through  
formula
\be
{\bf{a}} \; \; = \; \; {\bf{a}}' \; + \;{\bf{{\dot{\Omega}}}} \; \times \;
{\bf{r}}' \; + \; 2 \; {\bf{\Omega}} \; \times \; {\bf{v}}' \; + \; 
{\bf{\Omega}} \; \times \; ( {\bf{\Omega}} \times {\bf{r}}' ) \; \; \; \; , 
\label{5.11}
\ee
where $\; {\bf{r}}, \; {\bf{v}}, \; {\bf{a}} \;$ are the position, 
velocity and acceleration in the inertial frame, and  $\; {\bf{r}}', \; 
{\bf{v}}' \;$ and $\; {\bf{a}}' \;$ are those in the body frame. Here $\;
{\bf r}\,=\,{\bf r}'\;$. Mind though that $\;{\bf{v}}'\;$ and $\;{\bf{a}}'\;$ 
do not vanish in the body frame. They may be neglected on the same grounds as  
term $\;{\dot{I}}_i\Omega_i\;$ in (\ref{2.1}): precession of a body of 
dimensions $\;\sim\,{\it l}\;$, with period $\;\tau\;$, leads to 
deformation-inflicted velocities $\;v'\,\approx\;\epsilon\,\it{l}/\tau\;$ 
and accelerations $\;a'\,\approx\;\epsilon\,\it{l}/\tau^{2}\;$, $\;
\epsilon \;$ being the typical order of strains arising in the material. 
Clearly, for very small $\;\epsilon\;$, quantities $\; v' \;$ and $\; a'\;$ are
much less than the velocities and accelerations of the body as a whole (that 
are about $\;{\it{l}}/\tau\;$ and $\;{\it{l}}/{\tau}^2\;$, correspondingly). 
Neglecting these, we get, from (\ref{5.11}) and (\ref{5.2}), for the acceleration at 
point $\;(x,\,y,\,z)\;$:
\ba
\nonumber
{\bf{a}} \; \; = \; \; {{\bf{e}}_1} \; \left\{ \;
{\frac{1}{2}} \; {\Omega}_{\perp}^2 \; x \; \; \cos \; 2{\omega}t \; \; + \;\; 
{\frac{1}{2}} \; {\Omega}_{\perp}^2 \; y \; \; \sin \; 2{\omega}t \; \; + \; \;
z \; {\Omega}_{\perp} \; {\Omega}_3 \; h \; \; \cos \; {\omega}t  
\; \right\}\;+\\
\nonumber\\
\nonumber
+ \; \; {{\bf{e}}_2} \; \left\{
{\frac{1}{2}} \; {\Omega}_{\perp}^2 \; x \; \; \sin \; 2{\omega}t \; \; - \; \;
{\frac{1}{2}} \; {\Omega}_{\perp}^2 \; y \; \; \cos \; 2{\omega}t \; \; + \; \;
z \; {\Omega}_{\perp} \; {\Omega}_3 \; h \; \; \sin \; {\omega}t \; 
\; \right\}\;+\\  
\nonumber  \\
+ \; \; {{\bf{e}}_3} \; \left\{
{\Omega}_{\perp} \; {\Omega}_3 \; (2 \; - \; h) \; (x \; \; \cos \; {\omega}t 
\; \; + \; \; y \; \; \sin \; {\omega}t \; )
\; \right\}~~~.
\label{5.12}
\ea
Plugging this into (\ref{4.12}), with the proper boundary conditions imposed, 
yields, for an oblate prism of dimensions $\,2a\,\times\,2a\,\times\,2c\;$, 
$\;a\,>\,c\,$:
\be
\sigma_{xx} = \frac{\rho \Omega_\perp^2}{4}  
(x^2 - a^2) \; \cos 2 {\omega} t  \; \; , \; \; \; \;  
\sigma_{yy} = - \frac{\rho \Omega_\perp^2}{4} (y^2 - a^2) \;  
\cos 2 {\omega} t \; \; , \; \; \; \;  
\sigma_{zz} = 0 
\label{5.13}
\ee
\be
\sigma_{xy} \; \; = \; \; \frac{\rho}{4} \; \Omega_\perp^2 \; 
(x^2 \; \; + \; \; y^2 \; \; - \; \; 2a^2)
\; \; \sin \; 2 {\omega} t \; \; \; ,
\label{5.14}
\ee
\be
\sigma_{xz} \; \; = \; \; \frac{\rho}{2} \; {\Omega_\perp} \; {\Omega_3} \; 
\left[ \; h \; (z^2 \; - \; c^2)
\; \; + \; \; (2 \; - \; h) \; (x^2 \; - \; a^2) \; \right] \; \; 
\cos \; {\omega} t \; \; \; ,
\label{5.15}
\ee
\be
\sigma_{yz} \; \; = \; \; \frac{\rho}{2} \; {\Omega_\perp} \; {\Omega_3} \; 
\left[ \; h \; (z^2 \; - \; c^2)
 \; \; + \; \; (2 \; - \; h) \; (y^2 \; - \; a^2) \; \right] \; \; 
\sin \; {\omega} t \; \; \; .
\label{5.16}
\ee
In (\ref{5.12}) - (\ref{5.16}) we kept only time-dependent parts, because 
time-independent parts of the acceleration, stresses and strains are irrelevant
in the context of dissipation. A detailed derivation of (\ref{5.12}) - 
(\ref{5.16}) is presented in (Lazarian \& Efroimsky 1999). 

Formulae (\ref{5.13}) - (\ref{5.16})
implement the polynomial approximation to the stress tensor. This approximation
keeps the symmetry and obeys (\ref{4.12}) with (\ref{5.12}) plugged into it. The 
boundary condition are satisfied 
exactly for the diagonal components and only approximately for 
the off-diagonal components. The approximation considerably simplifies
calculations and entails only minor errors in the numerical factors in 
(\ref{5.21}). 

The second overtone 
emerges, along with the principal frequency $\,\omega\,$, in the expressions for stresses 
since the centripetal part of the acceleration is quadratic in $\; \bf{\Omega}
\;$. The kinetic energy of an oblate spinning body reads, according to
(\ref{2.8}), (\ref{5.3}), and (\ref{5.9}):
\begin{eqnarray}
T_{kin}\;=\;\frac{1}{2}\;\left[I\;\Omega_{\perp}^2\;+\;I_3\;{\Omega_3}^2\right]
\;=\;\frac{1}{2}\;\;\left[\;\frac{1}{I}\;\;\sin^2\;\theta\;\;+\;\; 
\frac{1}{I_3} \; \; \cos^2 \; \theta \; \right] \; J^2\;\;,\;\;\;
\label{5.17}
\end{eqnarray}
wherefrom
\be
\frac{dT_{kin}}{d\theta} \; \; = \; \; 
\frac{J^2}{I_3} \; (h \; - \; 1) \; \; \sin \; \theta \; \; \cos \; \theta 
\; \; =  \; \; \omega \; J \; \; \sin \; \theta \; \; \; .
\label{5.18}
\ee
The latter expression, together with (\ref{3.5}) and (\ref{4.7}), leads to:
\be
\frac{d\theta}{dt} \; = \; \left(\frac{dT_{kin}}{d\theta}\right)^{-1} 
\frac{dT_{kin}}{dt} \; = \; \left( \omega \; J \; \; \sin \; \theta 
\right)^{-1} \; \dot{W}\;\;,
\label{5.19}
\ee
where
\ba
\dot{W}\; = \; \dot{W}^{({\omega})}\; + \; \dot{W}^{(2{\omega})} \; = 
\; \omega \; \frac{W_0^{({\omega})}}{Q^{({\omega})}} \; + \; 2 \; \omega \; 
\frac{W_0^{({2\omega})}}{Q^{({2\omega})}}\;\approx 
\; \frac{2\,\omega}{Q} \; 
\left\{ <W^{({\omega})}> \; + \; 2\;< W^{({2\omega})}> \protect\right\}\;, 
\label{5.20}
\ea
the quality factor assumed to depend upon the frequency very 
weakly\footnote{The $\omega$-dependence of $\;Q\;$ should be taken into account
within frequency spans of several orders, but is irrelevant for frequencies 
differing by a factor of two.}. In the above formula, $\;W_0^{\omega}\;$ and 
$\;W_0^{2\omega}\;$ are amplitudes of elastic energies corresponding to the 
principal mode and the second harmonic. Quantities $<W^{\omega}>\,=\,W_0^{
\omega}/2\,$ and $\,<W^{2\omega}>\,=\,W_0^{2\omega}/2\,$ are the appropriate 
averages. Substitution of (\ref{5.13}) - (\ref{5.16}) into (\ref{4.7}), with 
further integration over the volume and plugging the result into (\ref{3.5}), 
will give us the final expression for the alignment rate:
\be
d \theta/dt \; = \; - \; \frac{3}{2^4} \; \sin^3 \theta \; \; \frac{63 \; 
(c/a)^4 \; \cot^2 \theta \; + \; 20}{[1+(c/a)^2]^4} \; \; \frac{a^2 \; 
\Omega^3_0 \; \rho}{\mu \; Q}
\label{5.21}
\ee
where 
\be
\Omega_0\;\equiv\;\frac{J}{I_3} \;\;\;\;
\label{5.22}
\ee
is a typical angular velocity. Deriving (\ref{5.21}), we took into account 
that, for an oblate $\;2a\,\times\,2a\,\times\,2c\;$ prism (where $\;a\,>\,c\;
$), the moment of inertia $\;I_3\;$ and the parameter $\;h\;$ read:
\be
I_3 \;  = \; \frac{16}{3} \; \rho \; a^4 \; c\;\;\;\;\;,\;\;\;\;\;\;\;\;
h \; \; \equiv \; \; \frac{I_3}{I} \; \; = \; \; \frac{2}{1 \; + \; (c/a)^2}  
\; \; \; \; \; . \; \; \; 
\label{5.23}
\ee
Details of derivation of (\ref{5.21}) are presented in (Lazarian \& Efroimsky 
1999)\footnote{Our expression (\ref{5.21}) presented here differs from the 
appropriate formula in Lazarian \& Efroimsky (1999) by a factor of 2, because 
in Lazarian \& Efroimsky (1999) we missed the coefficient 2 connecting 
$W_0^{(...)}$ with $W^{(...)}$.}. 

Formula (\ref{5.21}) shows that the major-inertia axis slows down its alignment
at small residual angles. For $\theta\rightarrow 0 $, the derivative ${\dot{\theta
}}\,$ becomes proportional to $\theta$, and thus, $\theta$ decreases 
exponentially slowly: $\theta = A \exp (-\zeta t)$, where $A$ and $\zeta$ are some 
positive numbers\footnote{This resembles the behaviour of a pendulum: if the pendulum 
is initially given exactly the amount of kinetic energy sufficient for the pendulum to 
move up and to point upwards at the end of its motion, then formally it takes an infinite 
time for the pendulum to stand on end.}. This feature, ''exponentially slow finish'', 
(which was also mentioned, with regard to the Chandler wobble, in Peale (1973), formula (55))
is natural for a relaxation process, and does not lead to an infinite relaxation time 
if one takes into account the finite resolution of the equipment. 
Below we shall discuss this topic at length. 

Another feature one might expect of (\ref{5.21}) would be a ``slow start'': 
it would be good if $\;d\theta/dt\;$ could vanish for $\;\theta\,\rightarrow\,\pi/2\;$. 
If this were so, it would mean that at $\;\theta\,=\,\pi/2\;$ (i.e., when the 
major-inertia axis is exactly perpendicular to the angular-momentum vector) the body 
``hesitates'' whether to start aligning its maximal-inertia axis along or opposite to 
the angular momentum, and the preferred direction is eventually determined by some stochastic 
influence from the outside, like (say) a collision with a small meteorite. This behaviour 
is the simplest example of the famous spontaneous symmetry breaking, and in this setting 
it is desirable simply for symmetry reasons: $\;\theta\,=\,\pi/2\;$ must be a position of an unstable 
equilibrium\footnote{Imagine a knife freely rotating about its longest dimension, and let the rotation
axis be vertical. This rotation mode is unstable, and the knife must eventually come to rotation about 
its shortest dimension, the blade being in the horizontal plane. One cannot say, though, which of the 
two faces of the blade will look upward and which downward. This situation is also illustrated by 
the pendulum mentioned 
in the previous footnote: when put upside down on its end, the pendulum ''hesitates'' in what 
direction to start falling, and the choice of direction will be dictated by some infinitesimally 
weak exterior interaction (like a sound, or trembling of the pivot, or an evanescent flow of air).}. Contrary to these 
expectations, though, (\ref{5.21}) leaves $\; d \theta /dt \;$ nonvanishing for $\;
\theta\,\rightarrow\,\pi/2\;$, bringing an illusion that the major axis  
leaves the position $\; \theta = \pi/2 \;$ at a finite rate. This failure of 
our formula (\ref{5.21}) comes from the inapplicability of our analysis in the
closemost vicinity of $\;\theta\,=\,\pi/2\;$.  This vicinity simply falls out 
of the adiabaticity realm adumbrated by (\ref{2.6}), because $\;\omega\;$ given
by (\ref{5.6}) vanishes for  $\;\theta\,\rightarrow\,\pi/2\,$ (then one can no longer
assume the relaxation to be much slower than the precession rate, and hence, the
averaging over period becomes illegitimate). 

One more situation, that does not satisfy the adiabaticity assertion, is when  
$\omega\,$ vanishes due to  $(h\,-\,1)\rightarrow 0$. This happens when $c/a$
approaches unity. According to (\ref{5.21}), it will appear that $d\theta /dt
$ remains nonvanishing for $c/a\rightarrow 1$, though on physical 
grounds the alignment rate must decay to zero because, for $c=a$, 
the body simply lacks a major-inertia axis.
 
All in all, (\ref{5.21}) works when $\theta$ is not too close to 
$\pi/2$ and $c/a$ is not too close to unity:
\be
-\;{\dot{\theta}}\;\ll\;(h\;-\;1)\;\frac{J}{I_3}\;\cos \theta \;=
\frac{1\;-\;(c/a)^2}{1\;+\;(c/a)^2}\;\Omega_0\;\cos \theta\;\;\;.
\label{5.24}
\ee
Knowledge of the alignment rate $\;\dot \theta\;$ as a function of the 
precession-cone half-angle $\;\theta\;$ enables one not only to write down a 
typical relaxation time but to calculate the entire dynamics of the process. In
particular, suppose the observer is capable of measuring the precession-cone  
half-angle $\;\theta\;$ with an error $\;\delta\;$. This observer will then 
compute, by means of (\ref{5.21}), the time needed for the body to change its 
residual half-angle from $\;\theta\;$ to $\;\theta\,-\,\Delta \theta\;$, for 
$\;\Delta\,\theta\,>\,\delta\;$. This time will then be compared with the results of 
his further measurements. Below we shall show that such observations will soon 
become possible for spacecraft.

First, let us find a typical relaxation time, i.e., a time span 
necessary for the major-inertia axis to shift considerably toward alignment 
with $\bf J\;$. This time may be defined as:
\begin{equation}
t_r \; \;  \; \; \equiv \;\; \int_{\theta_0}^{\delta} \; 
\frac{d \theta}{d \theta/dt} \; \; \; , \;\;\;
\label{5.25}
\end{equation}
$\theta_0$ being the initial half-angle of the precession cone ($\theta_0<\pi/2
$), and $\;\delta\;$ being the minimal experimentally-recognisable value of $\;
\theta\;$. A finite $\;\delta\;$ will prevent the ``slow-finish'' divergency. A
particular choice of $\;\theta_0\;$ and $\;\delta\;$ will lead to an 
appropriate numerical factor in the final expression for $\;t_r\;$. Fig.1 shows
that $\;t_r\;$ is not very sensitive to the choice of angle $\;\theta_0\;$, as 
long as this 
angle is not too small. This weak dependence upon the initial 
angle is natural since our approach accounts for the divergence at small angles
(``exponentially slow finish'') and ignores the ``slow start''. Therefore one can 
take, for a crude estimate, 
\be
\theta_0\,=\,\pi/2\;\;\;.\;\;
\label{5.26}
\ee
For $\;t_r\;$ it would give almost the same result as, say, $\;\pi/3\;$ or $\;
\pi/4\;$. A choice of $\;\delta\;$ must be determined exclusively by the 
accuracy of the observation technique: $\;\delta\;$ is such a minimally 
recognizable angle that precession within a cone of half-angle $\;\delta\;$ or 
less cannot be detected. Ground-based photometers measure the 
lightcurve-variation amplitude that is approximately proportional to the 
variation in the cross-sectional area of the wobbling body. In such sort of 
experiments the relative error is around {\it ~0.01}. In other words, only 
deviations from one revolution to the next exceeding {\it ~0.01 mag} may be 
considered real. This corresponds to precession-cone half-angles $\;\delta\,
\approx\,10^o\;$ or larger (Steven Ostro, private communication). Ground-based 
\break
\begin{figure}
\centerline{\epsfxsize=3.5in\epsfbox{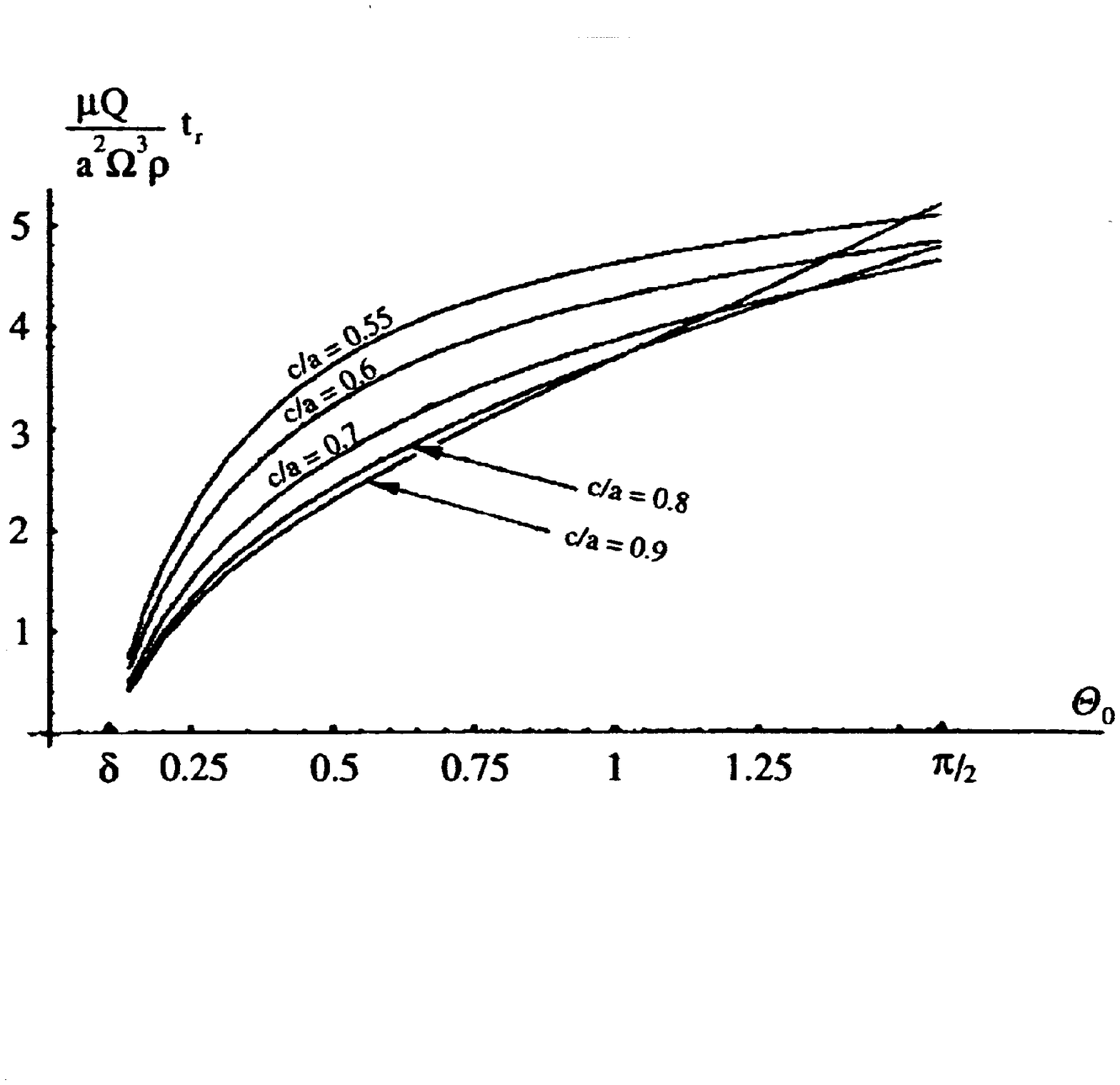}}
\bigskip
\caption{Precession of an oblate body: relaxation time $\,t_r\,$ as a 
function of 
$\,\theta_0\,$, where  $\,\theta_0\,$ is the initial value of the 
precession-cone half-angle $\,\theta\,$. Precession begins with the 
precession-cone half-angle $\,\theta\,$ equal to $\,\theta_0\,$, and 
effectively ends when  $\,\theta\,$ reaches the minimal measurable value $\;
\delta\;$.  On this plot, it is assumed that the initial value $\,\theta_0\,$ 
varies from $\;\delta\,=\,6^0\;$ through $\;\pi/2\;$. This choice of $\;\delta
\;$ corresponds to the current abilities of ground-based radars. 
(Spacecraft-based instruments provide $\;\delta\,=\,0.01^o\;.$)}
\end{figure}
\noindent\\
radars have a much sharper resolution and can grasp asteroid-shape details as 
fine as $\;10\;m\;$. This technique may reveal precession at half-angles of 
about 5 degrees. NEAR-type missions potentially may provide an accuracy of $
\,0.01^o\;$ (Miller et al. 1999). For a time being, we would lean to a 
conservative estimate
\be
\delta\;=\;6^o\;\;\;\;,\;\;\;
\label{5.27}
\ee
though we hope that within the coming years this limit may be reduced  
by three orders due to advances in the spacecraft-borne instruments.

Together, (\ref{5.21}), (\ref{5.25}) - (\ref{5.27}) yield dependences 
illustrated by Fig.1. Remarkably, $\;t_r\;$ is not particularly sensitive to 
the half-sizes' ratio $\;c/a\;$ when this ratio is between 0.5 - 0.9 (which is 
the case for realistic asteroids, comets and many spacecraft). Our formulae 
give:
\begin{eqnarray}
\nonumber
t_{(our \; result)} \; \approx \; (1\;-\;2) \; \frac{\mu \; Q}{
\rho \; 
a^2\; \Omega^3_0} \;\;\;\;\;\; for \;\;\;\;\;\; \theta_0\,\approx\,(2\;-\;3)\, \delta
\;=\;12\;-\;18^o\;\;\;\;\;\;;
\ea
\ba
t_{(our \; result)}\;\approx \;(3\;-\;4) \;\frac{\mu \; Q}{\rho 
\;a^2 \; \Omega^3_0} \; \; \; \; \;\;for \;\;\;\;\;\; \theta_0\,\approx\,\pi/4\;\;\;\;\;\;\;\;;\;\;\;\;
\label{5.28}
\ea
\ba
\nonumber
t_{(our \; result)} \; \approx \; (4\;-\;5) \; \frac{\mu \; Q}{
\rho \; a^2 \; \Omega^3_0} \;\;\;\;\;\; for \;\;\;\;\;\; \theta_0\,\stackrel{<}{\sim}\,\pi/2\;\;\;\;\;\;.\;\;\;\;
\end{eqnarray}
(Mind though that, according to (\ref{5.24}), $\theta$ should not approach 
$\pi/2$ too close.) To compare our results with a preceding 
study, recall that according to Burns \& Safronov (1973)
\begin{eqnarray}
t_{(B \; \& \;S)} \; \approx \; 100 \; \frac{\mu \; Q}{\rho \; a^2 \; 
\Omega^3_0} \; \; \; .\;\;\;
\label{5.29}
\end{eqnarray}
The numerical factor in Burns \& Safronov's formula is about $100$ for objects 
of small oblateness, i.e., for comets and for many asteroids. (For objects of 
irregular shapes Burns and Safronov suggested a factor of about $\,20$
in place of $\,100$.)

This numerical factor is the only difference between our formula and that of 
Burns \& Safronov. This difference, however, is quite considerable: for small
residual half-angles $\;\theta\;$, our value of the relaxation time is  
two orders smaller than that predicted by Burns \& Safronov. For larger 
residual half-angles, the times differ by a factor of several dozens. We see 
that the effectiveness of the inelastic relaxation was much underestimated by 
our predecessors. There are three reasons for this 
underestimation. The first reason is that our calculation was based on an
improved solution to the boundary-value problem for stresses. Expressions 
(\ref{5.13}) - (\ref{5.16}) show that an overwhelming share of the 
deformation (and, therefore, of the inelastic dissipation) is taking place in 
the depth of the body. This is very counterintuitive, because on a heuristic 
level the picture of precession would look like this: a centrifugal bulge, with
its associated strains, wobbles back and forth relative to the body as $\;
\Omega\;$ moves through the body during the precession period. This naive 
illustration would make one think that most of the dissipation is taking place 
in the shallow regions under and around the bulge. It turns out that in reality
most part of the deformation and dissipation takes place deep beneath the 
bulge (much like in the simple example with the liquid planet, that we provided 
in the end of section IV). The second, most important, reason for our formulae 
giving smaller 
values for the relaxation time is that we have taken into account the second 
harmonic. In many rotational states this harmonic turns to be a provider of the
main share of the entire effect. In the expression $\;(63 (c/a)^4\,\cot^2 
\theta + 20) \;$ that is a part of formula (\ref{5.21}), the term $\; 63 
(c/a)^4 \,\cot^2 \theta\;$ is due to the principal frequency, while the term $
\;20\;$ is due to the second harmonic\footnote{For calculational details, see 
Lazarian \& Efroimsky (1999).}. For $\;c/a\;$ belonging to the realistic 
interval $\;0.5\,- \,0.9\;$, the second harmonic contributes (after integration
from $\;\theta_0\;$ through $\;\delta\;$) a considerable input in the 
entire effect. This input will be of the leading order, provided the initial 
half-angle $\;\theta_0\;$ is not too small (not smaller than about $\,30^o\,$).
In the case of a small initial half-angle, the contribution of the second mode 
is irrelevant. Nevertheless it is the small-angle case where the discrepancy 
between our formula and (\ref{5.29}) becomes maximal. The estimate (\ref{5.29})
for the characteristic time of relaxation was obtained in Burns \& Safronov 
(1973) simply as a reciprocal to their estimate for $\;\dot \theta\;$; it 
ignores any dependence upon the initial angle, and thus gives too long 
times for small angles. The dependence of the dissipation rate of the values of
$\;\theta\;$ is the third of the reasons for our results being so different 
from the early estimate (\ref{5.29}).

Exploration of this, third, reason may give us an important handle on  
observation of asteroid relaxation. It follows from (\ref{5.21}) that a small 
decrease in the precession-cone half-angle, $\;-\,\Delta \theta\;$, will be 
performed during the following period of time:
\ba
\Delta t\;=\;(-\,\Delta \theta)\;\frac{2^4}{3}\;\frac{[1+(c/a)^2]^4}{63\;
(c/a)^4\;\cot^2 \theta \;+\;20}\;\frac{1}{\sin^3 \theta}\;\frac{\mu\;Q}{a^2\;
\Omega^3_0 \;\rho}\;\;\;.
\label{5.30}
\ea 
For asteroids composed of solid silicate rock, the density 
may be assumed $\;\rho\,\approx\,2.5\,\times\,10^3\;kg/m^3\;$, while the 
product in the numerator should be $\;\mu\,Q\,\approx\,1.5\,\times\,10^{13}\;
dyne/cm^2\,=\,1.5\,\times\,10^{12}\,Pa\;$ as explained in Efroimsky \& Lazarian
(2000). Burns \& Safronov suggested a much higher value of $\;3\,\times \,
10^{14}\;dyne/cm^2\,=\,3\,\times\,10^{13}\,Pa\;$, value acceptable within the  
terrestial seismology but, probably, inapplicable to asteroids.

For asteroids composed of friable materials, Harris (1994) suggests the 
following values: $\;\rho\,\approx\,2\,\times\,10^3\;kg/m^3\;$ and $\;\mu\,Q\,
\approx\,5\,\times\,10^{12}\;dyne/cm^2\,=\,5\,\times\,10^{11}\,Pa\;$. 
Naturally, this value is lower than those appropriate for solid rock (Efroimsky
\& Lazarian 2000), but in our opinion it is still too high for a friable 
medium. Harris borrowed the aforequoted value from preceding studies of Phobos
(Yoder 1992). Mind, though, that Phobos may consist not {\textit only} of rubble:
it may have a solid component in the centre. In this case, a purely rubble-pile
asteroid may have a lower $\;\mu\,Q\;$ than suggested by Harris. Anyway, as 
a very conservative estimate for a rubble-pile asteroid, we shall take the 
value suggested by Harris.

As for the geometry, let, for example, $\;\theta\,=\,\pi/3\;$ and $\;c/a\,=\,
0.6\;$. Then
\be
\Delta t\;=\;(-\,\Delta \theta)\;1.2\;\frac{\mu\;Q}{a^2\;\Omega^3_0 \;\rho}\;\;
\;.
\label{5.31}
\ee 
If we measure time $\;\Delta t\;$ in years, the revolution period $\;T\,=\,2\,
\pi/\Omega_0\;$ in hours, the maximal half-size $\;a\;$ in kilometers, and $
\;\theta\;$ in angular degrees ($\Delta \theta\,=\,\Delta \theta^o\,\times\,
1.75\,\times\,10^{-2}$), our formula (\ref{5.30}) will yield:
\be
\Delta t_{(years)}\;=\;(-\,\Delta \theta^o)\;\times\;1.31\;\times\;10^{-7}\;
\frac{\mu\;Q}{\rho}\;\frac{T^3_{(hours)}}{a^2_{(km)}}\;=\;0.33\;\frac{T^3_{(
hours)}}{a^2_{(km)}}\;\;\;,
\label{5.32}
\ee
where we accepted Harris' values of $\;\mu\,Q\,=\,5\,\times\,10^{11}\,Pa\;$ and
$\rho\,=\,2\,\times\,10^3\;kg/m^3\;$, and the angular resolution of 
spacecraft-based devices was assumed to be as sharp as $\;|\Delta \theta|\,=\,
0.01^o\;$, according to Miller et al. (1999).

\section{Triaxial and prolate rotators}

Typically, asteroids and comets have elongated shapes, and the above formulae 
derived for oblate bodies make a very crude approximation of the wobble of a 
triaxial or prolate body. In the case of a triaxial rotator, with $\;I_3 \geq I_2 
\geq I_1\;$, the solution to the Euler equations is expressable in terms of
elliptic functions. According to Jacobi (1882) and Legendre (1837), it will 
read, for $\;{\bf{J}}^2\; < \; 2\;I_2 \; T_{\small{kin}}\;$, as 
\begin{eqnarray}
\Omega_1\;=\;\gamma\;\,{\it{dn}}\left(\omega t , \; k^2 \protect\right)\;\;,
\;\;\;\;\Omega_2 \; = \; \beta \,\; \sn\left(\omega t , \; k^2 \protect\right)
\;\;,\;\;\;\;\Omega_3 \; = \; \alpha\;\,{\it{cn}}\left(\omega t,\;k^2
\protect\right)\;\;,\;\;\;
\label{1}
\end{eqnarray}
while for $\;\;{\bf{J}}^2\;>\;2\;I_2\;T_{\small{kin}}\;$ it will be: 
\begin{eqnarray}
\Omega_1\;=\;{\gamma}\;\,{\it{cn}}\left({\omega} t,\;{ k}^2
\protect\right)\;\;,\;\;\;\;\Omega_2\;=\;{\beta}\,\;{\it sn}\left({
\omega}t,\;{ k}^2 \protect\right)\;\;,\;\;\;\;\Omega_3\;=\;{\alpha
}\;\,{\it{dn}}\left({ \omega} t,\;{ k}^2 \protect\right)\;\;.\;\;\;
\label{2}
\end{eqnarray}
Here the precession rate $\,\omega\,$ and the parameters $\;\alpha ,\;\beta , \;
\gamma \,$ and $\;{k}\;$ are some algebraic functions of $\;I_{1,2,3}, \;
T_{\small {kin}}\;$ and $\;{\bf J}^2\;$. For example, $\;{k}\;$ is expressed by
\begin{eqnarray}
k\;=\;\sqrt{\frac{I_3-I_2}{I_2-I_1}\;\frac{{\bf J}^2-2I_1 T_{kin}}{2I_3 T_{kin}-{\bf J}^2}}\;\;\;\;\;,\;\;\;\;
\label{211}
\end{eqnarray}
for (\ref{1}), and by
\begin{eqnarray}
k\;=\;\sqrt{\frac{I_2-I_1}{I_3-I_2}\;\frac{2I_3 T_{kin}-{\bf J}^2}{{\bf J}^2-2I_1 T_{kin}}}\;\;\;\;\;,\;\;\;\;
\label{212}
\end{eqnarray}
for (\ref{2}). In the limit of oblate symmetry (when $\;I_2/I_1\,\rightarrow\,1\;$), solution
(\ref{2}) approaches (\ref{5.2}), while the applicability region of (\ref{1})
shrinks. Similarly, in the prolate-symmetry limit ($\,(I_3\,-\,I_2)/I_1\,
\rightarrow\,0\,$) the applicability realm of (\ref{2}) will become 
infinitesimally small. The easiest way of understanding this would be to 
consider, in the space $\;\Omega_1\;,\;\Omega_2\;,\;\Omega_3\;,$ the 
angular-momentum ellipsoid $\;\;{\bf{J}}^2\;=\;I_1^2\;\Omega_1^2\;+\;I_2^2\; 
\Omega_2^2 \; + \; I_3^2 \; \Omega_3^2 \;\;$. A trajectory described by the 
angular-velocity vector $\;\bf \Omega\;$ in the space $\;\Omega_1\;,\; \Omega_2
\;,\;\Omega_3\;$ will be given by a line along which this ellipsoid intersects 
the kinetic-energy ellipsoid $\;\;2\;T_{\small{kin}}\;=\;I_1\;\Omega_1^2 
\;+\;I_2\;\Omega_2^2\;+\;I_3\;\Omega_3^2\;\;,$ as on Fig.2. Through the 
relaxation process, 
the angular-momentum ellipsoid remains unchanged, while
the kinetic-energy ellipsoid evolves as the energy dissipates. Thus, the fast 
process, precession, will be illustrated by the (adiabatically) periodic motion
of $\;\bf \Omega\;$ along the line of ellipsoids' intersection; the slow 
process, relaxation, will be illustrated by the gradual shift of the moving vector
$\;\bf \Omega\;$ from one trajectory to another (Lamy \& Burns 1972). On Fig.2,
we present an angular-momentum ellipsoid for an almost prolate body whose 
angular momenta relate to one another as those of asteroid 433 Eros: $\;1\;
\times\;3\;\times\;3.05\;$ (Black et al. 1999). Suppose the initial energy was 
so high that $\;\bf\Omega\;$ was moving along some trajectory close to the pole
A on Fig.2. This pole corresponds to rotation of the body about its minor-inertia
axis. The trajectory described by $\;\bf \Omega\;$ about $\,A\,$ is almost circular 
and remains so until $\;\bf \Omega\;$ approaches the separatrix\footnote{This 
trajectory on Fig.2 being almost circular does not necessarily mean that the 
precession cone of the major-inertia axis about $\,\bf J\,$ is circular or almost 
circular.}. This process will be described by solution (\ref{1}). In the vicinity of 
separatrix, trajectories
will become noticeably distorted.
\break
\begin{figure}
\centerline{\epsfxsize=3.5in\epsfbox{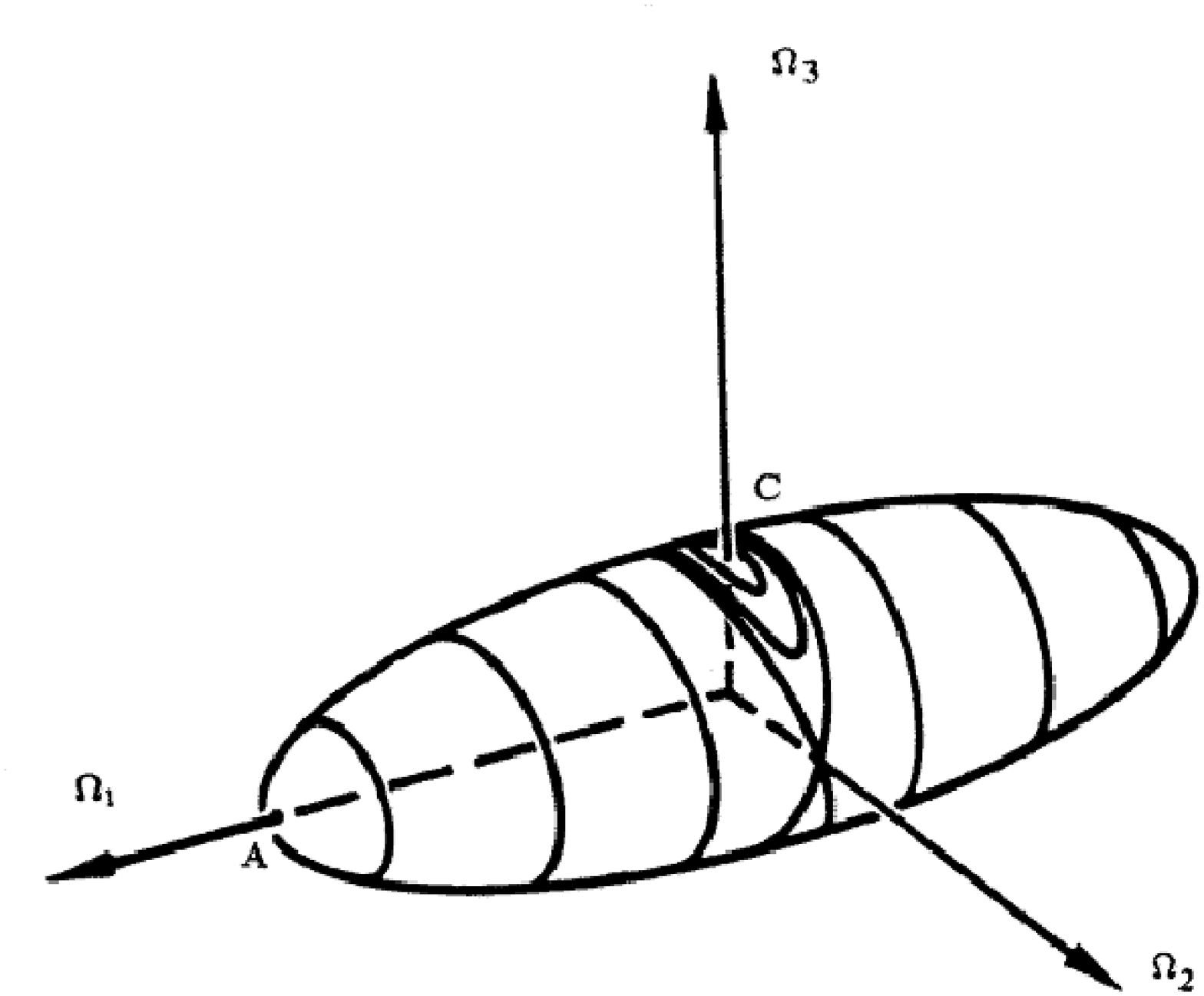}}
\bigskip
\caption{The constant-angular-momentum ellipsoid, in the angular-velocity space. The 
lines on its surface are its intersections with the kinetic-energy ellipsoids 
corresponding to different values of the rotational energy. The quasi-stable 
pole A is the maximal-energy configuration, i.e., the state when the body spins
about its minimal-inertia axis. The stable pole C symbolises the minimal-energy
state, i.e., rotation about the maximal-inertia axis. The angular-velocity 
vector precesses along the constant-energy lines, and at the same time slowly 
shifts from one line to another, approaching pole C. The picture illustrates 
the case of an elongated body: $I_3 \stackrel{>}{\sim}I_2>I_1$. The 
trajectories are circular near A and (in the case of an elongated body) remain 
virtually circular almost up to the separatrix. After the separatrix is crossed
(with chaotic flipovers possible), the body starts tumbling. The trajectories 
will regain a circular shape only in the closemost proximity of C.}
\end{figure}
\noindent\\
 Crossing of the separatrix may be accompanied
by stochastic flipovers\footnote{The flipovers are unavoidable if dissipation 
of the kinetic energy through one precession cycle is less than a typical 
energy of an occational interaction (a tidal-force-caused perturbation, for 
example).}. After the separatrix is crossed, librations will begin: $\;\bf 
\Omega\;$ will be describing not an almost circular cone but an elliptic one. This 
process will be governed by solution (\ref{2}). Eventually, in the closemost 
vicinity of pole C, the precession will again become almost circular. (This 
pole, though, will never be reached because the alignment of $\;{\bf{\Omega}}\;
$ towards $\;{\bf{J}}\;$ has a vanishing rate for small residual angles.) 
Parameter $\;k\;$ shows how far the tip of $\;\bf{\Omega}\;$ is 
from the separatrix on Fig.2: $\;k\;$ is zero in poles
A and C, and is unity on the separatrix. It is defined by (\ref{211}) when $\,\bf \Omega \,$ is between pole 
A and the separatrix, and by (\ref{212}) when $\,\bf \Omega \,$ is between the separatrix
and pole C. (For details see (Efroimsky 2000).)

If in the early stage of relaxation of an almost prolate ($I_3\,\approx\,I_2
$) body the tip of vector $\;\bf{\Omega}\;$ is near pole A, then its slow
departure away from A is governed by formula (9.22) in (Efroimsky 2000):
\ba
\nonumber
\frac{d\,\bra \sin^2 \theta \ket}{dt}\,=\\
\nonumber\\
\nonumber
-\,\frac{4\;\rho^2\;{\bf J}^2}{\mu\;Q(\omega)}\;\left(I_3\,-\,I_1\right)\;
\left(1\,-\;\bra \sin^2 \theta \ket \,\right)\;\left\{\,\omega\,S_1\,\left[2\;
\bra \sin^2 \theta \ket\;-\;1\;-\right.\right.\\
\nonumber\\
\nonumber
\left.\left.-\;\frac{1}{2}\;\frac{I_3\,-\,I_2}{I_2\,-\,I_1}\;\frac{I_1}{I_3}\;
\left(1\,-\;\bra \sin^2 \theta \ket \,\right)\right]\;-\right.\\
\nonumber\\
\left.-\;\omega\;S_0\;\frac{2\,I_1}{I_3}\;\left(1\,-\;\bra \sin^2\theta \ket \,
\right)\;+\;2\,\omega\,S_2\;\frac{Q(2\omega)}{Q(\omega)}\;\frac{2\,I_1}{I_3}\;
\left(1\,-\;\bra \sin^2 \theta \ket\,\right)\right\}\;.\;\;
\label{6.1}
\ea
where
\be
\omega\;=\;\sqrt{\frac{\left(2\;I_3\;T_{\small{kin}}\;-\;{\bf{J}}^2
\protect\right)\;\left(I_2\;-\;I_1\protect\right)}{I_1\;I_2\;I_3}}\;\approx\;
\frac{|{\bf{J}}|}{I_1}\;\sqrt{\frac{\left(I_3\;-\;I_1\protect\right)\;\left(I_2\;-\;I_1\protect\right)}{I_2\;I_3}}\;\sqrt{2\;\bra \sin^2 \theta \ket\;-\;1}\;\;,\;\;
\label{6.2}
\ee
$\theta\;$ is the angle between the angular-momentum vector $\bf J$ and the 
major-inertia axis $\bf{3}\;\,$; $\;S_{0,1,2}$ are some geometrical factors ($S_0\,=\,0\,$
in the case of $\,I_2\,=\,I_3\,$), and  $\;<...>\;$ symbolises an average over 
the precession cycle. For $\;<cos^2\theta>\;$ not exceeding $\;\approx\,1/7\;$, 
this equation has an exponentially decaying solution. For $\;c/a\,=\,0.6\;$ 
that solution will read:
\be
\Delta t\;\approx\;(-\,\Delta \bra {\theta } \ket )\;\times\;0.08\;\frac{\mu\;Q}{a^2\;
\Omega^3_0 \;\rho}\;\;\;.
\label{6.3}
\ee
Comparing this with (\ref{5.31}), we see that at this stage the relaxation
is about 15 times faster than in the case of an oblate body. 

During the later stage, when $\bf{\Omega}$ gets close to the 
separatrix, all the higher harmonics will come into play, and our estimate will
become invalid. How do the higher harmonics emerge? Plugging of (\ref{1}) or 
(\ref{2}) into (\ref{5.12}) will give an expression for the acceleration of an 
arbitrary point inside the body. Due to (\ref{4.12}), that expression will
yield formulae for the stresses. These formulae will be similar to 
(\ref{5.13} - \ref{5.16}), but will contain elliptic functions instead of the 
trigonometric functions. In order to plug these formulae for $\;\sigma_{ij}\;$ 
into (\ref{4.7}), they must first be squared and averaged over the precession 
cycle. For a rectangular prizm $\;2a\,\times \,2b\,\times \,2c\;$, a direct calculation
performed in (Efroimsky 2000) gives:
\be
\bra \sigma_{xx}^2 \ket = \frac{\rho^2}{4}\,(1-Q)^2\,\beta^4\,(x^2 - a^2)^2
\;\Xi_1 \; \; , \; \; \; \;  
\label{6.4}
\ee
\be
\bra \sigma_{yy}^2 \ket = \frac{\rho^2}{4}\,(S+Q)^2\,\beta^4\,(y^2 - b^2)^2\;
\Xi_1 \;\;\;,\;\;\;\;
\label{6.5}
\ee
\be
\bra \sigma_{zz}^2 \ket = \frac{\rho^2}{4}\,(1-S)^2\,\beta^4\,(z^2 - c^2)^2\;
\Xi_1 \;\;\;,\;\;\;\;
\label{6.6}
\ee
\be
\bra (Tr\,\sigma)^2 \ket =  \frac{\rho^2}{4}\,\beta^4\,\left\{(1-Q)(x^2 - 
a^2)^2\,+\,(S+Q)(y^2 - b^2)\;+\;(1-S)(z^2 - c^2)\right\}^2\;\Xi_1 \;\;\;,\;
\;\;\;
\label{6.7}
\ee
\be
\bra \sigma_{xy}^2 \ket = \frac{\rho^2}{4}\,\left\{(\beta \gamma + \alpha 
\omega k^2)(y^2 - b^2)+(\beta \gamma - \alpha \omega k^2)(x^2 - a^2)\right\}^2
\;\Xi_2 \;\;\;,\;\;\;\;
\label{6.8}
\ee
\be
\bra \sigma_{xz}^2 \ket = \frac{\rho^2}{4}\,\left\{(\beta \omega + \alpha 
\gamma)(z^2 - c^2)  + (\beta \omega - \alpha \gamma)(x^2 - a^2) \right\}^2\;
\Xi_3 \;\;\;,\;\;\;\;
\label{6.9}
\ee
\be
\bra \sigma_{yz}^2 \ket = \frac{\rho^2}{4}\,
\left\{(\alpha \beta + \omega \gamma)(z^2 - c^2)  + (\alpha \beta - \omega 
\gamma)(y^2 - b^2) \right\}^2\;\Xi_4 \;\;\;,\;\;\;\;
\label{6.10}
\ee
where $Q$ and $S$ are some combinations of $\,I_1, \,I_2,\,I_3\,$, defined by
formula (2.8) in (Efroimsky 2000). Factors $\,\Xi_{1,2,3,4}\,$ stand for 
averaged powers of the elliptic functions:
\ba
\nonumber
\Xi_1\;\equiv\;\bra\,\;\left(\,{\it sn}^2(u,\,k^2)\;-\;\,<\,{\it sn}^2(u,\,k^2)
\,>\,\;\right)^2\,\;\ket\;=\\
\nonumber\\
=\;< \,{\it sn}^4(u,\,k^2)\,> \;-\;< \,{\it sn}^2(u,\,k^2)\,>^2\;\;\;,\;\;\;
\label{6.11}
\ea
\ba
\nonumber
\Xi_2\;\equiv\; \bra \,\;\left( \, \sn (u,\,k^2)\; \cn (u,\,k^2)\;-\; \bra \,\;
\sn(u,\,k^2)\; \cn (u,\,k^2)\,\ket \,\right)^2\,\; \ket \;=\\
\nonumber\\
=\;<\,{\sn}^2(u,\,k^2)\;{\cn}^2(u,\,k^2)\,>\;-\;<\,\sn (u,\,k^2)\;\cn (u,\,k^2)
\,>^2\;\;\;,\;\;\;
\label{6.12}
\ea
\ba
\nonumber
\Xi_3\;\equiv\;\bra\,\;\left(\;{\it cn}(u,\,k^2)\;{\it dn}(u,\,k^2)\;-\;\bra\,\;
{\it cn}(u,\,k^2)\;{\it dn}(u,\,k^2)\,\;\ket\,\;\right)^2\,\;\ket\;=\\
\nonumber\\
=\;<\,{\it cn}^2(u,\,k^2)\;{\it dn}^2(u,\,k^2)\,>\;-\;<\,{\it cn}(u,\,k^2)\;
{\it dn}(u,\,k^2)\,>^2\;\;\;,\;\;\;
\label{6.13}
\ea
\ba
\nonumber\\
\nonumber
\Xi_4\;\equiv\;\bra\,\;\left(\,{\sn}(u,\,k^2)\;{\dn}(u,\,k^2)\;-\;\bra\,\;
{\sn}(u,\,k^2)\;{\dn}(u,\,k^2)\,\ket\,\;\right)^2\,\;\ket\;=\\
\nonumber\\
=\;<\,{\sn}^2(u,\,k^2)\;{\dn}^2(u,\,k^2)\,>\;-\;<\,{\sn}(u,\,k^2)\;{\dn}(u,\,k^2)\,>^2\;\;\;,\;\;\;
\nonumber\\
\label{6.14}
\ea
where averaging implies:
\ba
< ... > \;\equiv\;\frac{1}{\tau}\;\int_{0}^{\tau}\;\,.\,.\,.\,\;du\;\;\;,\;\;\;
\label{averaging}
\ea
$\tau\;$ being the mutual period of $\,\sn\,$ and $\,\cn\,$ and twice the period of $\;\dn\,$:
\be
\tau\;=\;4\;K(k^2)\;\equiv\;4\;\int_{0}^{\pi/2}\;
(1\,-\,k^2\,\sin^2 \psi )^{-1/2}\;d\,\psi\;\;\;.
\label{6.15}
\ee
The origin of expressions (\ref{6.11} - \ref{6.14}) can be traced from formulae
(8.4, ~8.6 - 8.13) in (Efroimsky 2000). For example, expression (\ref{5.11}), 
that gives acceleration of an arbitrary point inside the 
body, contains term $\;\sn^2(\omega t, \,k^2)\;$. (Indeed, one of the 
components of the angular velocity is proportional to $\,\sn (...)\,$, while 
the centripetal part of the acceleration is a quadratic form of the 
angular-velocity components.) The term $\;\sn^2(\omega t, \,k^2)\;$ in the 
formula for acceleration yields a similar term in the expression for $\;
\sigma_{xx}\;$. For this reason expression (8.6) in (Efroimsky 2000), that 
gives the {\bf{~time-dependent part}} of $\;\sigma_{xx}\;$, contains $\;\sn^2(.
..)\,-\,<sn^2(...)>\;$, wherefrom (\ref{6.11}) ensues.

Now imagine that in the formulae (\ref{6.4} - \ref{6.10}) the elliptic 
functions are presented by their series expansions over sines and cosines
(Abramovitz \& Stegun 1965):
\ba
{\sn}(\omega t,\,k^2)\;=\;\frac{2\pi}{k\,K}\sum_{n=1}^{\;\;\;\;\;\infty\;\;\;
*}\frac{q^{n/2}}{1\,-\,q^{n}}\;\sin\left(\omega_n\,t\right) \;\;\;\;
\;,\;\;\;\;\;
\label{6.16}
\ea
\ba
{\cn}(\omega t,\,k^2)\;=\;\frac{2\pi}{k\,K}\sum_{n=1}^{\;\;\;\;\;\infty\;\;\;
*}\frac{q^{n/2}}{1\,+\,q^{n}}\;\cos\left(\omega_n\,t\right)\;\;\;
\;\;,\;\;\;\;\;
\label{6.17}
\ea
\ba
{\dn}(\omega t,\,k^2)\;=\;\frac{\pi}{2\,K}\;+\;\frac{2\pi}{K}{\sum_{n=0}^{\;
\;\;\;\;\infty\;\;\;**}}\;\frac{q^{{n}/{2}}}{1\,+\,q^{n}}\;\cos\left(\omega_n
\, t\right) \;\;\;\;\;, \;\;\;\;\;
\label{6.18}
\ea
where 
\be
\omega_n\,=\,n\,\omega\,\pi/(2K(k^2))\;\;\;,\;\;\;q\,=\,
\exp(-\pi K({k'}^2)/K(k^2))\;\;\;,\;\;\;{k'\,}^2\,\equiv\,1\,-\,k^2\;\; 
\label{3}
\ee
and the function $\,K(k^2)\,$ is the complete elliptic integral of the first 
kind (see (\ref{6.15}) or (\ref{6.24})). A star in the superscript denotes a
sum over odd $n$'s only; a double star stands for a sum over even $n$'s. 
Plugging of 
(\ref{6.16}-\ref{6.18}) into (\ref{6.4}-\ref{6.10}) will produce, after 
squaring of $\;\sn , \, \cn , \, \dn\;$, an infinite amount of terms like $\;
\sin^2(\omega_n t)\;$ and $\;\cos^2(\omega_n t)\;$, along with an infinite 
amount of 
cross terms. The latter will be removed after averaging over the 
precession period, while the former will survive for all $n$'s and will average
to 1/2. Integration over the volume will then lead to an expression 
like (\ref{4.9}), with an infinite amount of contributions $\;\bra W_n \ket\;$ 
originating from all $\,\omega_n\,$'s, $\;n\,=\,1,\,.\,.\,.\,,\,\infty$. 
This is how an infinite amount of overtones comes into play. These overtones 
are multiples not of precession rate $\,\omega\,$ but of the "base frequency"
$\,\omega_1\,\equiv\,\omega \pi/(2K(k^2))\,$ which is lower than $\,\omega\,$.
Hence the stresses and strains contain not only Fourier components oscillating at  
frequencies higher than the precession rate, but also components oscillating at frequencies
lower than $\,\omega\,$. This is a very unusual and counterintuitive phenomenon.

The above series (called ''nome expansions'') typically converge very 
quickly, for $\,q\,\ll\,1\,$. Note, however, that $\,q\,\rightarrow\,1\,$ 
at the separatrix. Indeed, on approach to the separatrix we have: 
$\,k\,\rightarrow\,1\,$, wherefrom $\,K(k^2)\,\rightarrow\,\infty\,$; therefore 
$\,q\,\rightarrow\,1\,$ and  $\,\omega_n \,\rightarrow\,0\,$ (see eqn. 
(\ref{3})). The period of rotation (see (\ref{6.15})) becomes infinite. 
(This is the reason why near-separatrix states can mimic the principal one.)

Our previous work, Efroimsky (2000), addressed relaxation in the vicinity of 
poles. This case corresponds to $\;k\,\ll\,1\,$. For this reason we used, 
instead of (\ref{6.16} - \ref{6.18}), trivial approximations $\; \omega_1\,\approx \, \omega\; ,\;\sn (\omega 
t,\,k^2)\,\approx\,\sin (\omega t)\;,\;\cn (\omega t,\,k^2)\,\approx\,\cos (
\omega t)\;,\;\dn (\omega t,\,k^2)\,\approx\,$1. These approximations, along with
(\ref{6.4} - \ref{6.14}) enabled us to assume that the terms $\,\sigma_{xz}^2\,
$ and $\,\sigma_{yz}^2\,$ in (\ref{5.6}) are associated with the principal 
frequency $\,\omega\,$, while $\,<\sigma_{xx}^2>\,$, $\,<\sigma_{yy}^2>\,$, $\,
<\sigma_{zz}^2>\,$, $\,<(Tr\,\sigma )^2>\,$ and $\,\sigma_{xy}^2\,$ are 
associated with the second harmonic $\;2\,\omega\,$. No harmonics higher than
second appeared in that case. However, if we move away from the poles, 
parameter $\,k\,$ will no longer be small (and will be approaching unity as we 
approach the separatrix). Hence we shall have to take into account all terms in
(\ref{6.16} - \ref{6.18}) and, as a result, shall get an infinite amount of 
contributions from all $\;\omega_n\,$'s in (\ref{4.7} - \ref{4.9}). Thus we see
that the problem is very highly nonlinear. It is nonlinear even though the
properties of the material are assumed linear (strains $\;\epsilon\;$ are 
linear functions of stresses $\;\sigma\;$). Retrospectively, the nonlinearity 
originates because the dissipation rate (and, therefore, the relaxation rate) 
is proportional to the averaged (over the cycle) elastic energy stored in the 
body experiencing precession-caused alternating deformations. The average 
elastic energy is proportional to $\;<\sigma\,\epsilon>\;$, i.e., to  
$\;<\sigma^2>\;$. The stresses are proportional to the components of the 
acceleration, that are quadratic in the components of the angular velocity
(\ref{1} - \ref{2}). All in all, the relaxation rate is a quartic form of the 
angular-velocity components that are expressed by the elliptic functions $\;
(\ref{6.16} - \ref{6.18})\;$. 

A remarkable fact about this nonlinearity is that it produces oscillations of
stresses and strains not only at frequencies higher than the precession 
frequency $\;\omega\;$ but also at frequencies lower than $\,\omega$. This is 
evident from formula (\ref{3}): 
the closer we get to the separatrix (i.e., the closer $\,k^2\,$ gets to unity),
the smaller the factor $\;\pi/(2K)\;$, and the more lower-than-$\omega$ 
frequencies emerge.

A quantitative study of near-separatrix wobble will imply attributing extra 
factors of $\;\omega_n/Q(\omega_n)\;$ to each term of the series (\ref{4.9})
and investigating the behaviour of the resulting series (\ref{4.10}). This
study will become the topic of our next paper. Nevertheless, some qualitative 
judgement about the near-separatrix behaviour can be made even at this point.

For the calculation of the dissipation rate (\ref{4.10}), the value of the 
average elastic energy $\,<W>\,$ given by the sum (\ref{4.9}) is of no use
(unless each of its terms is multiplied by $\;\omega_n/Q (\omega_n)\;$ and 
plugged into (\ref{4.10})). For this reason, the values of the terms $\,<
\sigma_{ij}^2>\,$ entering (\ref{4.7}) are of no practical value either; only 
their expansions obtained by plugging (\ref{6.16} - \ref{6.18}) into (\ref{6.4}
 - \ref{6.14}) do matter. Nonetheless, let us evaluate $\,<W>\,$ near the 
separatrix. To that end, one has to calculate all $\,<\sigma_{ij}^2>\,$'s by 
evaluating (\ref{6.11} - \ref{6.14}). Direct integration in (\ref{6.11} - 
\ref{averaging}) leads to:
\ba
\Xi_1\;=\;\frac{1}{3\,k^4}\;\left\{k^2\;-\;1\;+\;\frac{2\,E}{K}\;\left(2\,-\,
k^2\right)\;-\;3\;\left(\frac{E}{K}\right)^2\right\}\;\;\;\;,\;\;\;
\label{6.19}
\ea
\ba
\Xi_2\;=\;\frac{1}{3\,k^4}\;\left\{2\;\left(k^2\;-\;1\right)\;+\;\frac{E}{K}\;
\left(\,-\,2\,-\,5\,k^2\right)\right\}\;\;\;\;,\;\;\;
\label{6.20}
\ea
\ba
\Xi_3\;=\;\frac{1}{3\,k^2}\;\left\{\frac{E}{K}\;\left(1\;+\;k^2\right)\;+\;
\left(k^2\;-\;1\right)\right\}\;\;\;\;,\;\;\;
\label{6.21}
\ea
\ba
\nonumber\\
\Xi_4\;=\;\frac{1}{3\,k^2}\;\left\{\frac{E}{K}\;\left(2\;k^2\;-\;1\right)\;+\;
\left(1\;-\;k^2\right)\right\}\;\;\;\;,\;\;\;
\label{6.22}
\ea
$K\,$ and $\;E\;$ being abbreviations for the complete elliptic integrals of the 1st and 2nd kind:
\ba
\nonumber
K\;\equiv\;K(k^2)\;\equiv\;\int_{0}^{\pi/2}\;
(1\,-\,k^2\,\sin^2 \psi )^{-1/2}\;d\,\psi\;\;\;,\\
\label{6.24}
\\
\nonumber
E\;\equiv\;E(k^2)\;\equiv\;\int_{0}^{\pi/2}\;
(1\,-\,k^2\,\sin^2 \psi )^{1/2}\;d\,\psi\;\;\;.\;\;\;
\ea
In the limit of $\,k\,\rightarrow\,1\,$, the expression for $\;K\;$ will 
diverge and all $\;\Xi_i\;$ will vanish. Then all $\,<\sigma_{ij}^2>\,$ will 
also become nil, and so will $\,<W>\,$. As all the inputs $\,<W(\omega_n)>\,$ 
in (\ref{4.10}) are nonnegative, each of them will vanish too. Hence the 
relaxation slows down near the separatrix. Moreover, it appears to completely 
halt on it. How trustworthy is this conclusion? On the one hand, it might have 
been guessed simply from looking at (\ref{6.15}): since for $\,k\,\rightarrow\,
1\,$ the period $\,4\,K(k^2)\,$ diverges (or, stated differently, since the 
frequencies $\,\omega_n\,$ in (\ref{3}) approach zero for each fixed n, then 
all the averages may vanish). On the other hand, though, the divergence of the 
period undermines the entire averaging procedure: for $\;\tau\;\rightarrow\;
\infty\;$, expression (\ref{3.3}) becomes pointless. Let us have a look at the 
expressions for the angular-velocity components near the separatrix. According 
to (Abramovits \& Stegun 1965), these expressions may be expanded into series 
over small parameter $\;(1-k^2)\;$:
\begin{eqnarray}
\nonumber
\Omega_1\;=\;\gamma\;\,{\it{dn}}\left(\omega t ,\; k^2 \protect\right)\;=\;
\gamma\;\,\left\{{\it{sech}}\left(\omega t \protect\right)^{\left.~ \right.}_{\left.~ \right.}\;+\right.\;\;\;\;\;\;\;\;\;\;\;\;\;\;\;\\
\nonumber\\
\left.+\;\frac{1}{4}\;(1\,-\,k^2)\;\left[{\mathit{sinh}}(\omega t)\;{\mathit
{cosh}}(\omega t)\;+\;\omega\,t\right]\;{\mathit{sech}}(\omega t)\;{\mathit{
tanh}}(\omega t) \right\}\;+\;O\left((1-k^2)^2\right)\;\;\;,\;\;
\label{6.224}
\end{eqnarray}
\ba
\nonumber
\Omega_2 \; = \; \beta \, \; {\sn} \left(\omega t , \; k^2 \protect\right)\;=
\;\beta\;\,\left\{{\mathit{tanh}}^{\left. \right.}_{\left.\right.}(\omega t)\;+\;\right.\;\;\;\;\;\;\;\;\;\;\\
\nonumber\\
\left.+\;\frac{1}{4}\;(1\,-\,k^2)\;\left[{\mathit{sinh}}(\omega t)\;{\mathit
{cosh}}(\omega t)\;-\;\omega\,t\right]\;{\mathit{sech}}^2(\omega t) \right\}\;+
\;O\left((1-k^2)^2\right)\;\;\;,\;\;
\label{6.25}
\end{eqnarray}
\ba
\Omega_3 \; = \; \alpha\;\,{\it{cn}}\left(\omega t,\;k^2\protect\right)\;=
\alpha\;\,\left\{{\mathit{sech}}(\omega t)\;-\;\right.\;\;\;\;\;\;\;\;\;\;\\
\nonumber\\
\left.-\;\frac{1}{4}\;(1\,-\,k^2)\;\left[{\mathit{sinh}}(\omega t)\;{\mathit
{cosh}}(\omega t)\;-\;\omega\,t\right]\;{\mathit{sech}}(\omega t)\;{\mathit{
tanh}}(\omega t) \right\}\;+\;O\left((1-k^2)^2\right)\;\;\;.\;\;
\label{6.26}
\end{eqnarray}
These expansions will remain valid for small $\;k^2\;$ up to the point $\;k^2\,
=\,1\;$, inclusively. It doesn't mean, however, that in these expansions we 
may take the limit of $\;t\,\rightarrow\,\infty\;$. (This difficulty arises
because this limit is not necessarily interchangeable with the infinite sum of 
terms in the above expansions.) Fortunately, though, for $\;k^2\,=\,1\;$, the 
limit expressions
\begin{eqnarray}
\Omega_1\;=\;\gamma\;\,{\it{dn}}\left(\omega t ,\; 1 \protect\right)\;=
\;\gamma\;\,{\it{sech}}\left(\omega t \protect\right)\;\;\;\;,\;\;
\label{6.27}
\end{eqnarray}
\ba
\Omega_2 \; = \; \beta \, \; {\sn} \left(\omega t , \; 1 \protect\right)\;=
\;\beta\;\,{\mathit{tanh}}(\omega t)\;\;\;\;,\;\;
\label{6.28}
\end{eqnarray}
\ba
\Omega_3 \; = \; \alpha\;\,{\it{cn}}\left(\omega t,\; 1 \protect\right)\;=\;
\alpha\;\,{\mathit{sech}}(\omega t)\;\;\;\;\;\;
\label{6.29}
\end{eqnarray}
make an {\underline{exact}} solution to (\ref{2.1}). 
Thence we can see what happens to vector $\;\bf{\Omega}\;$ when its tip  
is right on the separatrix. If there were no inelastic dissipation, the tip 
of vector $\;\bf{\Omega}\;$ would be slowing down while moving along the 
separatrix, and will come to halt at one of the middle-inertia homoclinic 
unstable poles (though it would formally take $\;\bf{\Omega}\;$ an infinite 
time to get there, because $\,\Omega_1\,$ and $\,\Omega_3\,$ will be 
approaching zero as $\;\sim\,\exp (-\omega t)\;\,$). When $\;\bf{\Omega}\;$ 
gets sufficiently close to the 
homoclinic point, the precession will slow down so that an observer
would get an impression that the body is in a simple-rotation state. In 
reality, some tiny dissipation will still be present even for very slowly 
evolving $\;\bf{\Omega}\;$. It will be present because this slow evolution will
cause slow changes in the stresses and strains. The dissipation will result in 
a further decrease of the kinetic energy, that will lead to a change in the 
value of $\;k^2\;$ (which is a function of energy; see (\ref{211}) and (\ref{212})). A 
deviation of $\;k^2\;$ away from unity will imply a shift of $\;\bf{\Omega}\;$ 
away from the separatrix towards pole C. So, the separatrix eventually 
{\underline{will}} be crossed, and the near-separatrix slowing-down does NOT 
mean a complete halt.

This phenomenon of near-separatrix slowing-down (that we shall call 
{\bf{lingering effect}}) is not new. In a slightly different context, it was 
mentioned by Chernous'ko (1968) who investigated free precession of a tank filled 
with viscous liquid and proved that, despite the apparent trap, the separatrix is
crossed within a finite time. Recently, the capability of near-intermediate-axis 
rotational states to mimic simple rotation was pointed out by Samarasinha, Mueller 
\& Belton (1999) with regard to comet Hale-Bopp. 

We, thus, see that the near-separatrix dissipational dynamics is very subtle, 
from the mathematical viewpoint. On the one hand, more of the higher overtones 
of the base frequency will become relevant (though the base frequency itself will 
become lower, approaching zero as the angular-velocity vector approaches the 
separatrix). On the other hand, the separartrix will act as a (temporary) trap,
and the duration of this lingering is yet to be estimated.

One should, though, always keep in mind that a relatively weak push can help the 
spinning body to cross the separatrix trap. So, for many rotators (at least, for 
the smallest ones, like cosmic-dust grains) the observational reality near separatrix 
will be defined not so much by the mathematical sophistries but rather by high-order 
physical effects: the solar wind, magnetic field effects, etc... In the case of a 
macroscopic rotator, a faint tidal interaction or a collision with a smaller body 
may help to cross the separatrix.

\section{Application to Asteroids and Comets}

Let us begin with 4179 Toutatis. This is an S-type asteroid analogous to stony 
irons or ordinary chondrites, so the solid-rock value of $\;\mu\,Q\;$ suggested
in Efroimsky \& Lazarian (2000) may be applicable to it:  $\;\mu\,Q\,\approx\,
1.5\,\times\,10^{13}\;dyne/cm^2\,=\,1.5\,\times\,10^{12}\,Pa\;$. Its density 
may be roughly estimated as $\;\rho\,=\,2.5\,\times\,10^3\;kg/m^3\;$ (Scheeres 
et al. 1998). Just as 
in the case of (\ref{5.32}), let us measure the time $\;\Delta t\;$ in years, 
the revolution period $\;T\;$ in hours ($T_{(hours)}\,=\,175$), the maximal 
half-radius $\;a\;$ in kilometers ($\;a_{(km)}\,=\,2.2\;$), and $\;\theta\;$ in
angular degrees ($|\Delta \theta^o|\,=\,0.01$). Then (\ref{6.3}) will yield:
\be
\Delta t_{(years)}\;\approx\;5.1\;\times\;10^{-2}\;\frac{T^3_{(hours)}}{a^2_{
(km)}}\;=\;5.6\;\times\;10^4\;years
\label{7.1}
\ee
Presently, the angular-velocity vector $\,\bf \Omega\,$ of Toutatis is at the 
stage of precession about $\,A\;$ (see Fig.2). However its motion does not obey the 
restriction $\;{\bra \cos^2 \theta  \ket}\,<\,1/7\,$ under which (\ref{6.3}) works 
well. A laborious calculation based on equations (2.16) and (A4) from Efroimsky (2000)
and on formulae (1), (2) and (11) from Scheeres et al (1998) shows that in the case of 
Toutatis $\;{\bra \cos^2 \theta  \ket}\,\approx\,2/7\,$. Since the violation is not 
that bad, one may still use (\ref{7.1}) as the zeroth approximation. Even if it is a 
two or three order of magnitude overestimate, we still see that the chances for 
experimental observation of Toutatis' relaxation are slim.
 
This does not mean, though, that one would not be able to observe asteroid 
relaxation at all. The relaxation rate is sensitive to the parameters of the body (size and 
density) and to its mechanical properties ($\,\mu Q\,$), but the precession period is certainly
the decisive factor.  Suppose that some asteroid is loosely-connected 
($\;\mu\,Q\,=\,5\,\times\,10^{12}\,dyne/cm^2\,=\,5\,\times\,10^{11}\,Pa\;$ and 
$\rho\,=\,2\,\times\,10^3\;kg/m^3\;$), has a maximal half-size 17 km, and is 
precessing with a period of 30 hours, and {\it{is not too close to the separatrix}}. 
Then an optical resolution of $|\Delta \theta^o|\,=\,0.01$ degrees will lead to 
the following time interval during which a $\;0.01^o\;$ change of the precession-cone 
half-angle will be measurable:
\be
\Delta t_{(years)}\;\approx\;2.12 \;\times\;10^{-2}\;\frac{T^3_{(hours)}}{a^2_{
(km)}}\;=\;2\;\;years\;\;\;\;
\label{7.2}
\ee
which looks most encouraging. In real life, though, it may be hard to observe precession
relaxation of an asteroid, for one simple reason: too few of them are in the states when
the relaxation rate is fast enough. Since the relaxation rate is much faster than believed 
previously, most excited rotators have already relaxed towards their principle states and 
are describing very narrow residual cones, too narrow to observe. The rare exceptions are 
asteroids caught in the near-separatrix ''trap''. These are mimicing the principal state. 

On these grounds, it is easy to guess the rotational state of 433 Eros: since it is not in a 
sweep-tumble mode, then most probably it is not precessing at all, or keeps an extremely narrow
residual cone. An almost circular precession with a half-angle of several 
degrees is very improbable because most likely it has already been transcended. Indeed, the 
observations have indicated no visible wobble (Yeomans $\it{et \; al}$ 2000).

What about comets? According to Peale and Lissauer (1989), for Halley's comet 
$\mu \approx 10^{10}\,dyne/cm^2\,=\,10^{9}\,Pa\,$ while $Q<100$, like for the regular 
ice. We are unsure if the values of order $\,100\,$ for $Q$ are acceptable; we would 
be more comfortable with values close to those of firn (heavy coarse-grained snow): $
Q\approx 1$. Then\footnote{Our 
estimate of $Q$ still remains rough, because the inner layers of the comet may contain 
amorphous water frost (Prialnik 1999), material whose attenuation 
properties may differ from those of firn.} $\mu Q \approx 10^{10}\,dyne/cm^2 \, = \, 10^{9}\,Pa$. As for the density of the cometary material, 
it is probable that the average density of a comet does not deviate much from $1.5\,\times
\,10^3\,kg/m^3$. Indeed, on 
the one hand, the major part of the material may have density close to that of 
firn, but on the other hand a typical comet will carry a lot of crust and dust
on and inside itself. Now, consider a comet of a maximal half-size 7.5 km (like
that of Halley comet (Houpis and Gombosi 1986)) precessing with a period of 3.7
days $\approx$ 89 hours (just as Halley does\footnote{Belton et al 1991, 
Samarasinha and A'Hearn 1991, Peale 1992}). If we once again 
assume the angular resolution of the spacecraft-based equipment to be  $|\Delta
\theta^o|\,=\,0.01$, it will lead us to the following damping time:
\be
\Delta t_{(years)}\;\approx\;5.65 \;\times\;10^{-5}\;\frac{T^3_{(hours)}}{a^2_{
(km)}}\;=\;0.7\;\;year\;\;\;.\;
\label{7.3}
\ee
This means that the cometary-relaxation damping may be measurable.

It also follows from (\ref{7.3}) that, to maintain the observed tumbling state 
of the Comet P/Halley, its jet activity should be sufficiently high\footnote{
The effect of outgassing upon the rotational state has been addressed in 
several articles. Wilhelm (1987) for the first time demonstrated numerically 
 that spin states can undergo significant changes due to outgassing 
torques. This was followed by Julian (1990). A detailed numerical 
treatment covering effects of outgassing over many orbits is presented in 
Samarasinha and Belton (1995).}.

\section{Application to Asteroid 433 Eros in Light of Recent Observations}

As already mentioned in the above section, asteroid 433 Eros is in a spin 
state that is either principal one or very close to it. This differs from
the scenario studied in (Black et al 1999). According to that scenario, an almost
prolate body would be spending most part of its history wobbling about the 
minimal-inertia axis. Such a scenario was suggested because the gap between the 
separatrices embracing pole C on Fig.2 is very narrow, for an almost prolate top, and
therefore, a very weak tidal interaction or impact would push the asteroid's angular 
velocity vector $\bf{\Omega}$ across the separatrix, away from pole C. This scenario
becomes even more viable due to the ''lingering effect'' described in section V, i.e.,
due to the relative slowing down of the relaxation in the closemost vicinity of the 
separatrix. 

Nevertheless, this scenario has not been followed by Eros. This could have happened for one
of the following reasons: either the dissipation rate in the asteroid is high 
enough to make Eros well relaxed after the recentmost disruption, or the asteroid simply has 
not experienced impacts or tidal interactions since times immemorial (since the early days of 
the Solar System, if we use the estimates by Burns \& Safronov (1973) who argued that the 
characteristic times of asteroid relaxation may be of order hundred of millions to billion years). 

The latter option is very unlikely: currently Eros is at the stage of leaving the main belt; it 
comes inside the orbit of Mars and approaches that of the Earth. It is then probable that Eros 
during its recent history was disturbed by the tidal forces that drove it out of the principal 
spin state.

Hence we have to prefer the former option, option that complies with our theory of precession 
relaxation. The fact that presently Eros is within less than 0.1 degree from its principal spin state 
means that the precession relaxation process is a very fast process, much faster than believed 
previously\footnote{Note that the complete (or almost complete) relaxation of Eros cannot be put down 
to the low values of the quality factor of a rubble pile, because this time we 
are dealing with a rigid monolith (Yeomans et al. 2000).}.

\section{Unresolved issues}

Our approach to calculation of the relaxation rate is not without its disadvantages. Some
of these are of mostly aesthetic nature, but at least one is quite alarming.

As was emphasised in the end of Section II, our theory is adiabatic, in that it assumes 
the presence of two different time scales or, stated differently, the superposition of two 
motions: slow and fast. Namely, we assumed that the relaxation rate is much slower than the 
body-frame-related precession rate $\,\omega\,$ (see formulae (\ref{2.5}) and (\ref{2.6})). 
This enabled us to conveniently substitute the dissipation rate by its average over a 
precession cycle. The adiabatic assertion is not necessarily fulfilled when $\,\omega\,$ 
itself becomes small. This happens, for example, when the dynamical oblateness 
of an oblate ($\,I_3\,>\,I_2\,=\,I_1\,\equiv\,I\,$) body is approaching zero:
\be
(h\,-\,1)\,\rightarrow\,0\;\;\;,\;\;\;\;\;\;\;\;\;\;\;h\;\equiv\;I_3/I\;\;\;\;.\;\;\;
\label{14.1}
\ee
Since in the oblate case $\,\omega\,$ is proportional to the oblateness (see (5.4)), it too 
will approach zero, making our adiabatic calculation inapplicable. This is the 
reason why one cannot and shouldn't compare our results, in the limit of $\,(h\,-\,1)\,\rightarrow 
\,0\,$, with the results obtained by Peale (1973) for an almost-spherical oblate body. 

In the general, triaxial case, our result, should not be compared, in the limit of weak 
triaxiality, to those presented in Peale, Cassen \& Reynolds (1979) and Yoder (1982), 
because those papers addressed not free dissipation but {\it{tidal}} dissipation. Our results, 
in the limit of weak triaxiality,  should not be compared either to those obtained by Yoder \& 
Ward (1979) for Venusian wobble-damping rate. The results of Yoder \& Ward (1979) are correct 
in the limit they were designed for, i.e., for an almost spherical planet. None of the asteroids 
and comets are almost spherical; hence they are subject to our approach, not to that 
of Yoder \& Ward. 

Another minor issue, that has a lot of mathematics in it but hardly bears any physical significance, 
is our polynomial approximation (\ref{5.13} - \ref{5.16} $\,$,$\,$ \ref{6.4} - \ref{6.10}) to the stress 
tensor. As explained in Section V, this approximation keeps the symmetry $\,\sigma_{ij}\,=\,\sigma_{ji}
\,$ and exactly satisfies (\ref{4.12}) with (\ref{5.12}) plugged in. The boundary conditions are fulfilled 
exactly for the diagonal components of the tensor and approximately for the off-diagonal elements. In the
calculation of the relaxation rate, this approximation will result in some numerical factor, and it is 
highly improbable that this factor differs much from unity.

A more serious difficulty of our theory is that it cannot, without further refinement, give a reasonable estimate
for the duration of the near-separatrix slowing-down mentioned in the end of Section VI. On the one hand,
many (formally, infinitely many) overtones of the base frequency $\,\omega_1\,$ come into play near the
separatrix; on the other hand, the base frequency approaches zero. Thence, it will take some extra 
work to account for the dissipation associated with the stresses oscillating at $\,\omega_1\,$ and with its 
lowest overtones. (The dissipation due to the stresses at these low frequency cannot be averaged over their 
periods.)  

There exists, however, one more, primary difficulty of our theory. Even though our calculation predicts a 
much faster relaxation rate than believed previously, it still may fail to account for the observed 
relaxation which seems to be even faster than we expect. This paper was already in press when Andrew Cheng 
confirmed the preliminary conclusion of the NEAR team, that the upper limit on non-principal axis rotation 
is better than 0.1 angular degree\footnote{Andrew Cheng, personal communication.}. How to interpret such a 
tough observational limit on Eros' residual precession-cone width? Our theory does predict very swift relaxation, but it also shows that the relaxation slows down near the separatrix and, especially, in the closemost vicinity 
of poles $\,A\,$ and $\,C\,$. Having arrived to the close vicinity of pole $\,C\,$, the angular-velocity 
vector $\,\Omega\,$ must exponentially slow down its further approach to $\,C\,$ (see the paragraph after 
equation (5.23)). For this reason, a body that is monolithic (so that its $\,\mu Q\,$ is not too low) and whose motion is sometimes influenced by tidal or other interactions,  must demonstrate to us at least some narrow residual precession cone. As already mentioned, for the past million or several millions of years Eros has been at the stage of leaving the main belt. It comes inside the Mars orbit and approaches the Earth. It is possible that 
Eros experienced a tidal interaction within the said period of its history. Nevertheless it
is presently in or extremely close to its principal spin state. The abscence of a visible residual 
precession not only disproves the old theory but also indicates that our new-born theory, too, may 
be incomplete. In particular, our $\,Q$-factor-based empirical description of attenuation should become the
fair target for criticisms, because it ignores several important physical effects.

One such effect is material fatigue. It shows itself whenever a rigid material 
is subject to repetitive load. In the case of a wobbling asteroid or comet, the stresses are tiny, 
but the amount of repetitive cycles, accumulated over years, is huge. At each cycle, the picture of emerging stresses is virtually the same. Moreover, beside the periodic stresses, there exists a constant component 
of stress. This may lead to creation of ''weak 
points'' in the material, points that eventually give birth to cracks or other defects. This may 
also lead to creep, even in very rigid materials. The creep will absorb some of the excessive energy
associated with precession and will slightly alter the shape of the body. The alteration will be such
that the spin state becomes closer to the one of minimal energy. It will be achieved through the slight 
change in the direction of the principal axes in the body. If this shape alteration is due to the emergence 
of a considerable crack or displacement, then the subsequent damping of precession will be performed by a finite step, not gradually. 

Another potentially relevant phenomenon is the effect that a periodic forcing (such as the solar gravity gradient) would have on the evolution and relaxation of the precession dynamics. It is possible that this 
sort of forcing could influence the precessional dynamics of the body\footnote{I am thankful to Daniel Scheeres who drew my attention to this effect.}.

\section{Rubble heap versus monolith}

Above we mentioned one of the most important discoveries of the NEAR-Schoemaker mission: Eros 
is a well-connected monolith. This brings up an interesting issue that is still unresolved.

At present, most astronomers lean toward the rubble-pile hypothesis, in regard to both asteroids and 
comets. The hypothesis originated in mid-sixties (${\ddot{O}}$pik 1966) and became a dominating theory 
in the end of the past century (Burns 1975; Asphaug \& Benz 1994; Harris 1996; Asphaug \& Benz 1996; 
Bottke \& Melosh 1996a,b; Richardson, Bottke \& Love 1998; Bottke 1998, Bottke, Richardson \& Love 1998; Bottke, Richardson, Michel \& Love 1999, Pravec \& Harris 2000). 

Sometimes comets get rent apart by the tidal forces (Asphaug \& Benz 1996, Sekanina 1982, Melosh \& Schenk 1993). On these and other grounds many researchers conclude that
all comets are weakly connected. A possible counter argument may be the following: since the
comets, when warmed up by the Sun, are prone to tidal disintegration, then perhaps,
the weakest comets have already perished and only the strongest have survived. Hopefully,
our understanding of the subject will improve after the Deep Impact mission reaches its goal. Meanwhile, 
we would lean towards the moderate viewpoint (Efroimsky \& Lazarian 2000): {\it{at least some 
of the comets are loosely connected conglomerates, but we do not know if all or even if most of them are
like that}}

In the case of asteroids, it may be unwise of us to completely reject the rubble-pile hypothesis. This hypothesis rests on
several strong arguments the main of which is this: the large fast-rotating asteroids are near the rotational breakup limit for aggregates with no tensile strength. Still, we would object to two of the arguments often used in support this theory. One such 
dubious argument is the low density of asteroid 253 Mathilde. The low density of
Mathilde (Veverka et al 1998, Yeomans et al 1998) may indeed evidence of high porosity. However, in our opinion, the word ''porous'' is not necessarily a synonim to ''rubble-pile'', even though in the 
astronomical community they are often used as synonims. In fact, a material may have high porosity and, at the same time, be rigid. 

Another popular argument, that we would contest, is the one about crator shapes. Many colleagues believe
that a rigid body would be shattered into smitherines by collisions; therefrom they infer that the asteroids must be soft, i.e., rubble. In our opinion, though, a rigid but highly porous 
material may stand very energetic collisions without being destroyed, because its porous structure damps 
the impact. 

Finally, it is know from the construction engineering that some materials, initially friable, become relatively rigid after being heated up (like, for example, asphalt).  They remain porous and may be prone to creep, but they are, nevertheless, sufficently rigid and well connected.

For these three reasons, we expressed in Efroimsky \& Lazarian (2000) our conservative opinion on the subject:
{\it{at least some asteroids are well-connected solid chunks, though we are uncertain whether this is true for all asteroids.}} This opinion met a cold reaction from the community. However, it is
supported by the recentmost findings. The monolithic nature of Eros is the most important of these.
Other include 1998KY26 studied in 1999 by Steven Ostro and his team: from the radar and optical observations,
the team inferred that this body, as well as several other objects, is monolithic (Ostro et al 1999). 

Still, we have to admit that the main argument in favour of rubble-pile hypothesis (the absence of large fast rotators) remains
valid.

\section{conclusions}

1. In many spin states, dissipation at frequencies different from the
   precession frequency makes a major input into the inelastic-relaxation 
   process. These frequencies are overtones of some "basic" frequency, that is 
   LOWER than the precession frequency. Thereby we encounter a very unusual example
   of nonlinearity: the principal frequency (precession rate) gives birth not only to 
   higher frequencies but also to lower frequencies.

2. Distribution of stresses and strains over the volume of a 
   precessing body is such that a major share of inelastic 
   dissipation is taking place deep inside the body, not in its shallow 
   regions, as thought previously. These and other reasons make inelastic 
   relaxation far more effective than believed hitherto.

3. However, if the rotation states that are close to the separatrix on Fig.2,  
   the lingering effect takes place: both precession and precession-damping 
   processes slow down. Such states (especially those close to the homoclinic point) 
   may mimic the principal rotation state.  

4. A finite resolution of radar-generated images puts a limit on our ability of
   recognising whether an object is precessing or not. Relaxation-caused 
   changes of the precession-cone half-angle may be observed. 
   Our estimates show that the modern spacecraft-based instruments are well fit
   for observations of the asteroid and cometary wobble relaxation. In many
   rotation states, relaxation may be registered within relatively short 
   periods of time (about a year).

5. Measurements of the damping rate will provide us with valuable 
   information on attenuation in small bodies, as well as on
   their recent histories of impacts and tidal interactions

6. Since inelastic relaxation is far more effective than presumed earlier, 
   the number of asteroids expected to wobble with a recognisable half-angle of
   the precession cone must be lower than expected. (We mean the predictions 
   suggested in (Harris 1994).) Besides, some of the small bodies may be in the
   near-separatrix states: due to the afore mentioned lingering effect, these 
   rotators may be ``pretending'' to be in a simple rotation state.

7. Even though our theory predicts a much higher relaxation rate than believed previously,  
   this high rate may still be not high enough to match the experimentally available data. 
   In the closemost vicinity of the principal spin state the relaxation rate must decrease
   and the rotator must demonstrate the "exponentially-slow finish". Asteroid 433 Eros is 
   a consolidated rotator whose $Q$-factor should not be too low. It is possible that this asteroid 
   was disturbed sometimes in its recent history by the tidal forces. Nevertheless, 
   it shows no visible residual precession. Hence, there may be a possibility that we shall have 
   to seek even more effective mechanisms of relaxation. One such mechanism may be creep-caused
   deformation leading to a subsequent change of the position of the principal axes in the body.

\section{What is to be done.}

Our further advance in the theoretical analysis of the phenomenon and in
planning the appropriate missions should include several steps. 

1. Our previous work (Efroimsky 2000)
   accounts for the dynamics at the stage when the angular-velocity vector $\;
   \bf{\Omega}\;$ and the major-inertia axis of the body describe almost 
   circular cones about the angular-momentum vector; that corresponds to $\;
   \bf{\Omega}\;$ describing almost circular trajectories on Fig.2. The next
   step would be to get an expression for the damping rate of a wobbling 
   triaxial rotator at the other stages of precession. In particular, it would be important to
   estimate the duration of the near-separatrix lingering, i.e., the time during
   which a rotator can mimic a simple rotation state.

2. Second, it is important to improve the precision of our calculation by 
   taking into account the real shapes of precessing bodies: it would be more 
   natural to model a body not by a prism but by an ellipsoid. This will demand
   a more refined mathematical approach to the appropriate boundary-value 
   problem for stresses.

3. Instrumentation on spacecraft have angular resolution of 0.01 degree (0.6 
   arcmin) or even better. It is a separate data-handling problem to make this 
   resolution translate into a similar resolution in the precession-cone 
   half-angle $\theta$. (Or, in the general case of a triaxial or prolate 
   rotator, into a similar resolution in the averaged-over-a-cycle $\;\sin^2
   \theta\;$.) 

4. Last, and by no means least, the mechanical and physical properties of 
   the asteroid and cometary materials must be studied. Much work in this 
   direction has already been done (Klinger et al. 1996; Muinonen \& Lagerros 
   1998; Remo 1994, 1999; Prialnik 1999), but our knowledge of attenuation in 
   small bodies still remains very basic, and consists more of hypotheses than 
   of facts. So we are in a bad need of both experimental and theoretical 
   results on attenuation in the materials asteroids and comets are made of. At the same time, it 
   is the future measurements of the relaxation rate that will shed light on 
   the material properties of the tumbling objects.

5. The above program, if carried out, will open up realistic perspectives for  
   measuring the wobbling bodies' relaxation rate. Observation of spin states 
   is naturally a part of any rendezvous mission. 
   In future, though, it would be better to perform not one but a series 
   of such measurements, by each such mission. In the case of comets, it would be 
   good to measure the spin state shortly before the
   perihelion (about 3 AU from the Sun or farther, i.e., before the outgassing 
   of water begins). The second measurement should be performed shortly after 
   the perihelion. Finally, at least one more observation would be in order 
   several months (or, even better, years) later. Such a scheme  
   would show the dynamics of both excitation and damping. 
   Calculations show that wobble-damping measurements have a good chance 
   of success. In the case of an asteroid, a success of such an experiment will
   crucially depend upon the structure of object: it may be difficult to 
   observe damping of a solid-rock asteroid, because the 
   dissipation in solid rock is slow. In rubble-pile asteroids 
   dissipation is several orders faster, and we may have a good chance of 
   observing relaxation of such rotators. In the case of a comet, we have a very good chance
   to register precession relaxation within a less than year period of time, if the spin state is
   not on the separatrix.\\

{\bf Acknowledgements}  

I am grateful to A. Erikson, A.Lazarian, M.Levi, S.Mottola, S.Ostro, G.Ryabova, N.Samarasinha, V.Sidorenko, 
D.Scheeres and other colleagues who so kindly spent their time discussing with me 
the issues raised in this article. I am especially thankful to J.Burns and S.Peale 
who suggested a lot of useful corrections while reviewing the early version of this 
paper. I am also grateful to B.Marsden and I.Shapiro for the support and encouragement 
they provided at the start of this project. Finally, my truly special thanks go to the 
reviewer, William Newman, whose comprehensive report on the manuscript helped me to 
considerably improve it and to enrich it with several illustrative examples.  

\pagebreak

\end{document}